\documentclass[pdflatex,sn-mathphys-ay]{sn-jnl}% Math and Physical Sciences Author Year Reference Style
\usepackage{graphicx}%
\usepackage{multirow}%
\usepackage{amsmath,amssymb,amsfonts}%
\usepackage{amsthm}%
\usepackage{mathrsfs}%
\usepackage[title]{appendix}%
\usepackage{xcolor}%
\usepackage{textcomp}%
\usepackage{manyfoot}%
\usepackage{booktabs}%
\usepackage{algorithm}%
\usepackage{algorithmicx}%
\usepackage{algpseudocode}%
\usepackage{listings}%
\usepackage{subcaption}
\usepackage{placeins}
\usepackage{multirow}
\usepackage{natbib}
\usepackage{hyperref}
\usepackage{subeqnarray}
\usepackage{soul}

\begin{document}

\title[Article Title]{Perpendicular rod wake/aerofoil interaction: microphone array and TR-PIV insights via SPOD and beamforming analysis}


\author*[1,2]{\fnm{Filipe} \sur{Ramos do Amaral}}\email{filipefra@ita.br}

\author[1]{\fnm{Marios Ioannis} \sur{Spiropoulos}}\email{marios-ioannis.spiropoulos@ensma.fr}

\author[1]{\fnm{Florent} \sur{Margnat}}\email{florent.margnat@univ-poitiers.fr}

\author[1]{\fnm{David} \sur{Marx}}\email{david.marx@univ-poitiers.fr}

\author[1]{\fnm{Vincent} \sur{Valeau}}\email{vincent.valeau@univ-poitiers.fr}

\author[1]{\fnm{Peter} \sur{Jordan}}\email{peter.jordan@univ-poitiers.fr}

\affil[1]{\orgdiv{DFTC-2AT}, \orgname{Institut Pprime, CNRS -- Université de Poitiers -- ISAE-ENSMA, UPR 3346}, \orgaddress{\street{11 Boulevard Marie et Pierre Curie}, \city{Poitiers}, \postcode{86073}, \state{Vienne}, \country{France}}}

\affil[2]{\orgdiv{Divisão de Aeronáutica}, \orgname{Instituto Tecnológico de Aeronáutica}, \orgaddress{\street{Praça Marechal Eduardo Gomes, 50}, \city{São José dos Campos}, \postcode{12228-900}, \state{São Paulo}, \country{Brasil}}}

%%%%%%%%%%%%%%%%%%%%%%%%%%%%%%%%%%%%%%%%%%%%%%
% Abstract
%%%%%%%%%%%%%%%%%%%%%%%%%%%%%%%%%%%%%%%%%%%%%%

\abstract{This paper investigates the acoustic and velocity fields due to a circular rod and an aerofoil placed in the wake of, and perpendicular to, a rod.
Simultaneous measurements were conducted using a microphone array and time-resolved particle image velocimetry (TR-PIV).
The interaction was characterized through acoustic spectra and the coherence between microphone signals and the three velocity components.
Coherent structures were identified with Spectral Proper Orthogonal Decomposition (SPOD) using a norm  based either on turbulence kinetic energy (SPOD-u) or on pressure (SPOD-p).
An advantage of SPOD-p is that it identifies velocity modes associated with a large acoustic energy.
Peaks of energy were observed at $\mathit{St} \approx 0.2$ and $0.4$--Strouhal numbers based on rod diameter and free-stream velocity.
At $\mathit{St} \approx 0.2$, the dominant feature is von Kármán vortex shedding from the rod.
At $\mathit{St} \approx 0.4$, a wave-train structure in the rod wake impinging on the aerofoil leading edge is captured by the rank-1 SPOD-p mode, with coherence levels reaching 60\% for the $u_2$ component (upwash/downwash relative to the aerofoil).
This structure also appears at $\mathit{St} \approx 0.2$, but as the rank-2 SPOD-p mode.
A mode-switching occurs around $\mathit{St} \approx 0.3$: below this value, the rank-1 mode corresponds to von Kármán shedding (cylinder branch), while above it, the rank-1 mode tracks the interaction of the aerofoil with the rod wake (aerofoil branch). 
Both branches were also identified via beamforming using low-rank cross-spectral matrices derived from SPOD-p modes.}

%%%%%%%%%%%%%%%%%%%%%%%%%%%%%%%%%%%%%%%%%%%%%%
% Key-words
%%%%%%%%%%%%%%%%%%%%%%%%%%%%%%%%%%%%%%%%%%%%%%


\maketitle

%%%%%%%%%%%%%%%%%%%%%%%%%%%%%%%%%%%%%%%%%%%%%%
\section{Introduction}
\label{sec:introduction}
%%%%%%%%%%%%%%%%%%%%%%%%%%%%%%%%%%%%%%%%%%%%%%

% Motivation
Noise due to turbulence and impinging vortices on an aerofoil leading edge appears in many practical problems and industrial applications.
Examples include rotor wings, such as the interaction between atmospheric turbulence and wind turbine blades~\citep{liu2017review}, the interaction of a blade wake with subsequent blades~\citep{brooks2004blade, roger2014vortex, quaglia20173d, raposo2024turbulence}, helicopter rotors~\citep{yung2000rotor, coton2004helicopter, thurman2023blade}, turbo-machinery~\citep{moreau2024turbomachinery}, ingestion of turbulence by a fan~\citep{wang2021computational}, unsteady gusts impinging on wing profiles~\citep{schlinker1983rotor, hales2023mathematical}, and interactions between propeller wakes and wings~\citep{felli2021underlying}, as well as aircraft landing-gear/flap interactions~\citep{oerlemans2004experimental, khorrami2014aeroacoustic, khorrami2015assessment, pottpollenske2017study, zhao2020noise}, among others.
The present paper focuses on the interaction between a rod wake and an aerofoil profile.

% Vortex-aerofoil
The simplest way to model rod wake/aerofoil interaction is by considering a vortex impinging on an aerofoil.
In order to study blade-vortex interaction (BVI) noise, experiments were conducted by~\citet{schlinker1983rotor} and~\citet{ahmadi1986experimental}.
They considered a tip vortex, generated by an upstream aerofoil, impinging vertically on a helicopter rotor blade model and proposed simplified models to predict the acoustic noise.
Later,~\citet{howe1988contributions} presented a general model for the sound produced when a vorticity field is intersected by a rigid aerofoil, modelled as acoustically compact and with a large aspect ratio (infinite span) in low-Mach-number flow.
\citet{roger2014vortex} and~\citet{quaglia20173d} extended this by considering an aerofoil of finite span, enhancing the previous models for more realistic configurations.
The numerical simulations by~\citet{zehner2018aeroacoustic} focused on the interaction of an isolated vortex impinging on a rotating blade.
On the other hand,~\citet{spiropoulos2024analysis},~\citep{spiropoulos2025aeroacoustics} studied the interaction between a vortex and a wedge using Howe’s model of a point vortex interacting with a semi-infinite half-plane.
The BVI noise modelling studies cited above employed various vortex formulations, including modifications in their orientation with respect to the aerofoil, to assess their impact on the sound field.
Overall, the authors found that the upwash/downwash velocity component, linked to unsteady lift on the aerofoil, is ultimately responsible for sound generation.

% Rod-aerofoil
The rod wake/aerofoil configuration consists of a cylinder placed upstream of an aerofoil.
Regarding the cylinder's orientation, it is typically in tandem with the aerofoil, i.e., its spanwise direction is aligned with that of the aerofoil.
For sufficiently high Reynolds numbers, a von Kármán vortex street develops in the rod/cylinder wake, which then impinges on the aerofoil~\citep{jacob2005rod}.
%The spectral signature shows broad peaks at a Strouhal number of approximately 0.2 (based on the rod diameter $d$ and free-stream speed $U_\infty$) and its harmonics, as well as a broadband component.
%The first harmonic of the shedding frequency, at a Strouhal number of approximately 0.4, is associated with cylinder drag fluctuations~\citep{giret2015noise}.
The largest time-scale of the wake corresponds to a Strouhal number of approximately 0.2 (based on the rod diameter $d$ and free-stream speed $U_\infty$), and is associated with the shear-layer flapping periodic motion, which is also that of lift fluctuation.
During this period, two vortices of opposite sign are shed into the cylinder wake.
Consequently, the drag is fluctuating at twice that frequency, namely $\mathrm{St} \approx 0.4$.
This is the frequency at which the aerofoil is impinged by vortices, disregarding their signs.
Consistently with the standard terminology, $\mathrm{St} \approx 0.2$ hereafter refers to as the vortex shedding frequency.
Peak broadening occurs as a result of the interaction between the von Kármán vortices and multiple turbulent scales, while the broadband component is related to the flow turbulence, which decays more gradually at higher frequencies~\citep{boudet2005wake}.

\citet{casalino2003prediction} conducted far-field acoustic and rod wall pressure measurements, as well as unsteady Reynolds-averaged Navier--Stokes (URANS) computations on a rod/aerofoil tandem configuration at low Mach numbers.
Only a single set-up was tested: constant flow speed (and Mach number), rod diameter, aerofoil chord, and the distance between the rod and aerofoil.
Noise levels were significantly enhanced by the presence of the aerofoil compared to rod-only results.
Good agreement was observed between experimental measurements and sound field computations using the Ffowcs-Williams Hawkings (FW--H) analogy, especially when three-dimensional spanwise effects were considered.
Later,~\citet{jacob2005rod} extended the work by~\citet{casalino2003prediction}, also performing particle image velocimetry (PIV) and hot-wire anemometry (HWA) measurements, together with complementary large-eddy simulation (LES) computations.
A parametric study was conducted, varying the rod diameter and flow speed (and Mach number).
For the rod-only acoustic measurements, the authors observed broad peaks around the vortex shedding frequency and its harmonics -- a feature that was amplified when considering the rod/aerofoil configuration.
The vortex shedding frequency was slightly shifted to lower values in the presence of the aerofoil, suggesting a weak feedback mechanism between the aerofoil and the rod.
Additionally, the vortex shedding peak spectral levels scaled with the Mach number to the power of 5.2, implying that the aerofoil does not behave as a compact dipole source (as assumed by~\citet{howe1988contributions}, for example), which would predict a power law of 6.
The 5.2 Mach power law is consistent with the diffraction of turbulent eddies by edges, suggesting that the sound source is located at the aerofoil edges.
This finding is corroborated by the PIV experiments, which show that the region around the aerofoil leading edge contains the dominant sound source, especially when analysing two-point velocity correlations, surpassing the rod and aerofoil trailing-edge regions.
Proper orthogonal decomposition (POD)~\citep{berkooz1993proper}, combined with vortex identification algorithms, was applied to the PIV database~\citep{jacob2005rod} and corresponding LES and URANS computations~\citep{boudet2005wake}.
The flow exhibits low-rank behaviour, with the first two modes dominating the flow dynamics, and excellent agreement was observed between the PIV and LES results~\citep{boudet2005wake}.
These modes correspond to the convection of von Kármán vortices toward the aerofoil leading edge and their interaction.
The experimental database by~\citet{jacob2005rod} was further explored by~\citet{greschner2008prediction}, who investigated the role of volume terms in the FW--H analogy using detached-eddy simulation (DES), which improved agreement between experimental and numerical acoustic spectra, particularly in the high-frequency range.

Several other authors have used~\citet{jacob2005rod}'s database as a benchmark to test their computational aeroacoustics (CAA) methods.
Among them,~\citet{berland2010numerical} and~\citet{berland2011parametric} implemented a direct noise calculation (DNC) method that does not require sound source modelling.
They achieved good agreement between numerical and experimental results, although some overestimation of spectral levels at certain Strouhal numbers was noted, likely due to differences in the signal lengths between simulations and experiments.
\citet{berland2011parametric} conducted a parametric study modifying the distance $D$ between the rod and the aerofoil, noting that the radiated sound field was most intense in the range $4 \lesssim D/d \lesssim 10$.
Shifts in the vortex shedding frequency as a function of $D/d$ were also observed.
\citet{jiang2015numerical} performed a similar parametric study, varying the distance between the cylinder and aerofoil in their LES+FW--H simulations.
They identified two flow regimes: one that suppresses the von Kármán street for $D/d \leq 2$ (short distances) and another where the von Kármán street persists for $D/d \geq 4$ (long distances), leading to vortex shedding.
The first regime, in which vortex impingement is weakened, produced lower noise levels compared to the second, which exhibited broadband and tonal noise at the shedding frequency and its harmonics.
\citet{eltaweel2011numerical} conducted LES computations using a boundary-element method (BEM) to calculate the acoustic radiation.
Their findings aligned with previous studies~\citep{jacob2005rod, boudet2005wake, greschner2008prediction}, showing that rod wake interaction with the aerofoil leading edge dominates sound generation, especially at low frequencies.
Trailing-edge noise becomes significant at higher frequencies.
Directivity plots reveal a clear dipole pattern in the normal direction at the vortex shedding frequency.
For higher frequencies, the directivity becomes more complex, with multiple lobes caused by scattering at the aerofoil leading and trailing edges and the rod.
\citet{giret2015noise} also performed LES+FW--H computations and noted constructive and destructive interferences between the rod and the aerofoil, depending on the frequency and observation angle.
Considering 0 and 180 degrees as angles aligned with the aerofoil leading and trailing edges, respectively,~\citet{giret2015noise} found that for angles below 45 degrees and above 135 degrees, the rod/aerofoil spectral levels at the vortex shedding frequency were lower than those of the aerofoil alone, indicating destructive interference.
At 90 degrees, the opposite occurred, with constructive interference.
The directivity plots by~\citet{giret2015noise} show a clear, nearly compact dipole at the vortex shedding frequency ($\mathit{St} \approx 0.2$), with maximum radiation perpendicular to the aerofoil chord.
At double the vortex shedding frequency ($\mathit{St} \approx 0.4$), a dipolar pattern emerges, with maximum radiation at 0 and 180 degrees, i.e., aligned with the aerofoil chord.

% Summary of the noise physics
In summary, studies on BVI and tandem rod/aerofoil interaction noise reveal that:
(i) the upwash/downwash velocity component is linked to sound generation;
(ii) the aerofoil leading edge contains the dominant noise source;
(iii) the aerofoil does not behave as a compact source;
(iv) two flow regimes exist depending on the rod/aerofoil distance, i.e., one that suppresses the von Kármán vortex street and presents broadband noise spectra (short distances), and another that develops the von Kármán vortex street, presenting broadband and tonal noise at the shedding frequency and its harmonics (long distances);
(v) at the vortex shedding frequency, the source directivity resembles a compact dipole, with maximum intensity levels perpendicular to the aerofoil chord, while the first harmonic of the vortex shedding frequency, associated with cylinder drag, exhibits a dipolar pattern with maximum intensity aligned with the aerofoil chord.
It is important to note that these features were observed for a rod configuration in tandem with the aerofoil.

% Objectives
As discussed in the previous paragraphs, most of the literature on rod wake/aerofoil interaction focuses on a rod positioned in tandem with the aerofoil, except for~\citet{spiropoulos2025aeroacoustics}, who modelled the problem considering a vortex impinging vertically on the aerofoil.
Additionally, much attention has been focused on identifying the contributions of both the rod and aerofoil to the total sound field.
Regarding flow structure characterization, only a few studies have touched on this subject~\citep{jacob2005rod, boudet2005wake, giret2015noise}.
Furthermore, there is a lack of time-resolved acoustic and velocity field databases that are simultaneously measured.
To address this gap, the present paper aims to characterize the flow structures that contribute to sound generation in the interaction between a perpendicular rod and an aerofoil profile, i.e., the rod is perpendicular to the aerofoil.
An extensive experimental campaign was conducted, including time-resolved PIV (TR-PIV) and microphone array measurements.
Two set-ups were explored with TR-PIV: one with the field of view perpendicular to the aerofoil plane, i.e., along the rod span, and the other along the aerofoil plane, i.e., perpendicular to the rod.
To post-process the data, snapshot version of spectral POD (SPOD)~\citep{towne2018spectral, schmidt2020guide} algorithm was employed to reveal the most energetic structures for both velocity and pressure fields.

% Paper organization
The remainder of the paper is organised as follows.
\S~\ref{sec:methodology} addresses the methodology, including the experimental set-up and procedures, as well as details on data post-processing.
\S~\ref{sec:results} presents the results, where the acoustic and TR-PIV databases are explored through energy spectra, coherence levels, coherent structures extracted with SPOD and acoustic maps obtained with conventional beamforming.
Finally, \S~\ref{sec:conclusions} provides concluding remarks.

%%%%%%%%%%%%%%%%%%%%%%%%%%%%%%%%%%%%%%%%%%%%%%
\section{Methodology}
\label{sec:methodology}
%%%%%%%%%%%%%%%%%%%%%%%%%%%%%%%%%%%%%%%%%%%%%%

%%%%%%%%%%%%%%%%%%%%%%%%%%%%%%%%%%%%%%%%%%%%%%
\subsection{Experimental set-up}
\label{sec:methodology_set-up}

The experiments were conducted at the Institut Pprime anechoic facility BETI (\emph{Bruit Environnement Transport Ingénierie}) in Poitiers, France.
The open-section wind tunnel has a 0.7~$\times$~0.7~m$^2$ cross-sectional area and is 1.5~m long.
The turbulence level is below 0.5\% at 50~m/s.
The anechoic chamber has a volume of 90~m$^3$ and a cut-off frequency of 200~Hz.

\begin{figure}[h]
	\centering
	\begin{subfigure}{0.45\textwidth}
		\includegraphics[width=\textwidth]{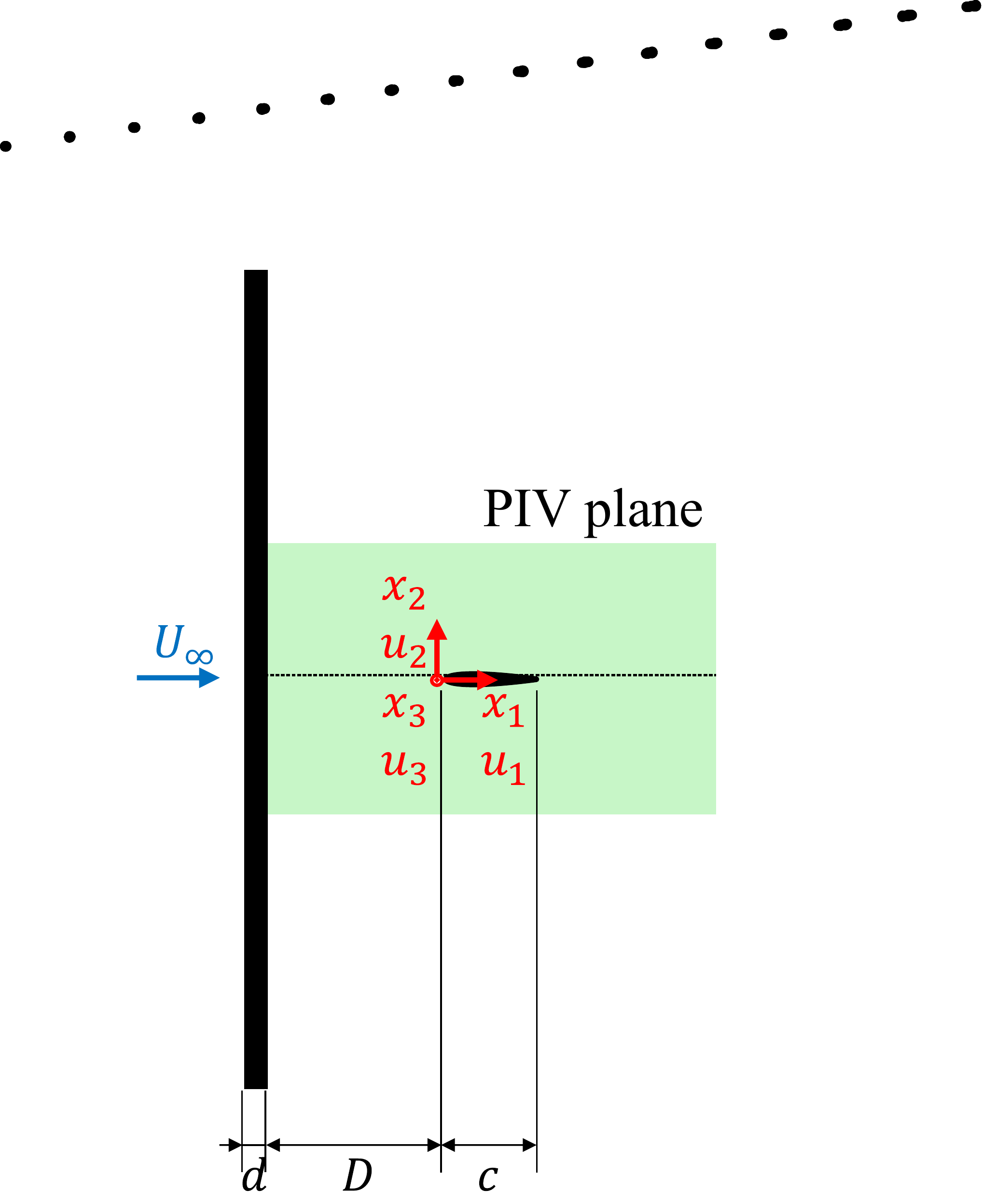}
		\caption{Lateral view (vertical plane, $x_1$--$x_2$)}
		\label{fig:sketch_vertical}
	\end{subfigure}
	\begin{subfigure}{0.45\textwidth}
		\includegraphics[width=\textwidth]{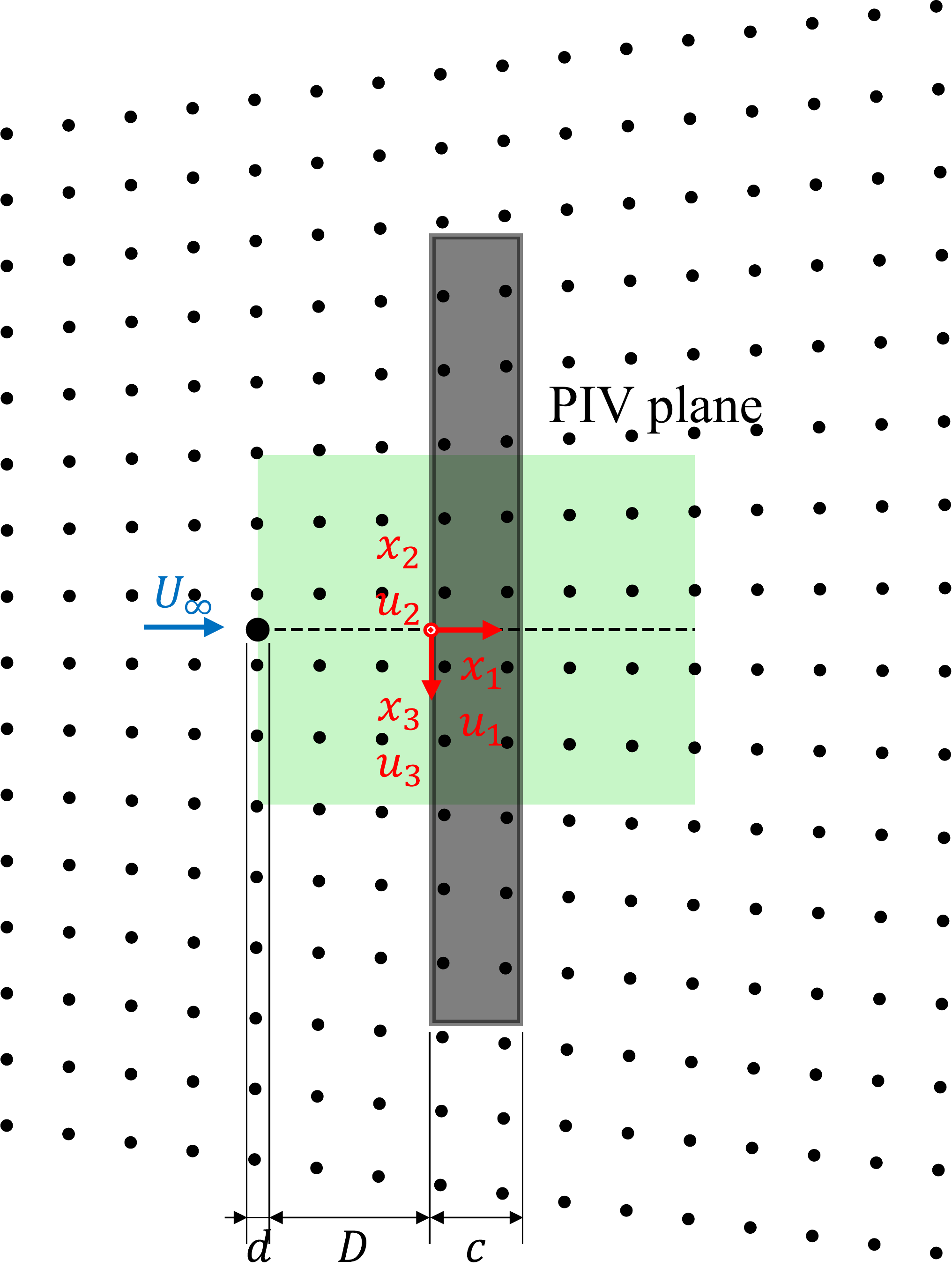}
		\caption{Top view (horizontal plane, $x_1$--$x_3$)}
		\label{fig:sketch_horizontal}
	\end{subfigure}
	\caption{Sketch representing the TR-PIV and acoustic experiment set-up, including the rod (black rectangle and circle), aerofoil (grey profile and rectangle), microphone array (black dots), TR-PIV field of view (green rectangle), coordinate system (in red), and key dimensions. The dotted black line in both frames indicates the position of the TR-PIV laser sheet, which is perpendicular to the shown view. The flow direction is from left to right.}
	\label{fig:sketch_PIV_acoustics}
\end{figure}

Figure~\ref{fig:sketch_PIV_acoustics} provides lateral and top view sketches of the experimental set-up used in the study.
A vertical cylinder/rod with a diameter of $d = 20$~mm and length of $s_\mathrm{rod} = 900$~mm was positioned $l = 310$~mm downstream from the open-section convergent exit.
Downstream of the rod, a NACA0012 aerofoil with a chord length of $c = 100$~mm ($c/d = 5$) and span of $s_\mathrm{aer} = 890$~mm was mounted perpendicular to the rod's axis, with an incidence angle of $\alpha = 0^\circ$.
The distance between the rod and the aerofoil was set to $D = 200$~mm ($D/d = 10$).

The coordinate system origin is located at the aerofoil leading edge at mid-span, and the axes are indicated in red in the figure.
The coordinates $x_1$, $x_2$, and $x_3$ represent the streamwise, upwash/downwash (cylinder span), and aerofoil span directions, respectively, as shown in Figure~\ref{fig:sketch_PIV_acoustics}.

The free-stream velocity was set to $U_\infty = 30$~m/s, which corresponds to $\mathit{Re}_c \approx 200{,}000$ for the aerofoil in air at 20$^\circ$C.
At this flow condition, tonal trailing-edge noise is expected~\citep{paterson1973vortex, arbey1983noise}, but it is suppressed here by forcing boundary-layer transition using a sandpaper strip placed at 5\% chord from the leading edge on both the suction and pressure sides of the aerofoil.
The type of sandpaper is P80, with an average grain size of 200 $\mu$m.

For the cylinder, the Reynolds number is $\mathit{Re}_d \approx 40{,}000$, corresponding to the subcritical regime.
At this value, the spanwise coherence length at the lift fluctuation frequency is approximately five diameters, with no significant dependence on Reynolds number~\citep{margnat2023cylinder}.
At the specified velocity, the acoustic wavelength associated with the lift fluctuation frequency is approximately 57 cylinder diameters, 11 aerofoil chords, and 6 rod--aerofoil distances.
For the drag fluctuation frequency, these values are halved.
This indicates that the entire interaction region is acoustically compact at the dominant frequencies.

Two-dimensional, three-component (2D-3C) time-resolved particle image velocimetry (TR-PIV) experiments, based on LaVision hardware, were conducted in both the vertical (Figure~\ref{fig:sketch_vertical}) and horizontal (Figure~\ref{fig:sketch_horizontal}) planes using two different set-ups.
For the vertical configuration (lateral view, $x_1$--$x_2$, Figure~\ref{fig:sketch_vertical}), the laser sheet was positioned at the aerofoil mid-span (dotted black line in Figure~\ref{fig:sketch_horizontal}).
For the horizontal configuration (top view, $x_1$--$x_3$, Figure~\ref{fig:sketch_horizontal}), the laser sheet was positioned 3~mm above the aerofoil's thickest point (dotted black line in Figure~\ref{fig:sketch_vertical}).

Two Nd:YAG lasers--one Continuum MESA PIV with a nominal power of 2 $\times$ 9 mJ at 152 $\mu$s and one Photonics Industries DMX 150-532 DH with a nominal power of 2 $\times$ 15 mJ at 152 $\mu$s, both operating at a wavelength of 532 nm--were aligned and synchronized to illuminate the entire field of view.
The lasers were synchronized using the modules LaVision HighSpeed Controller PTU (Continuum MESA PIV) and R\&D Vision EG (DMX 150-532 DH).
Moreover, synchronization among the lasers and the microphones array was performed using the Q1 signal from the MESA laser, which was employed to trigger the acoustic recordings.
The lasers synchronization was measured using a photodiode.
The laser power supply system, together with cooling system, were moved outside the anechoic chamber to ensure that the main sources of parasitic noise do not corrupt the acoustic measurements.

\begin{figure}[h]
	\centering
	\includegraphics[width=0.8\textwidth]{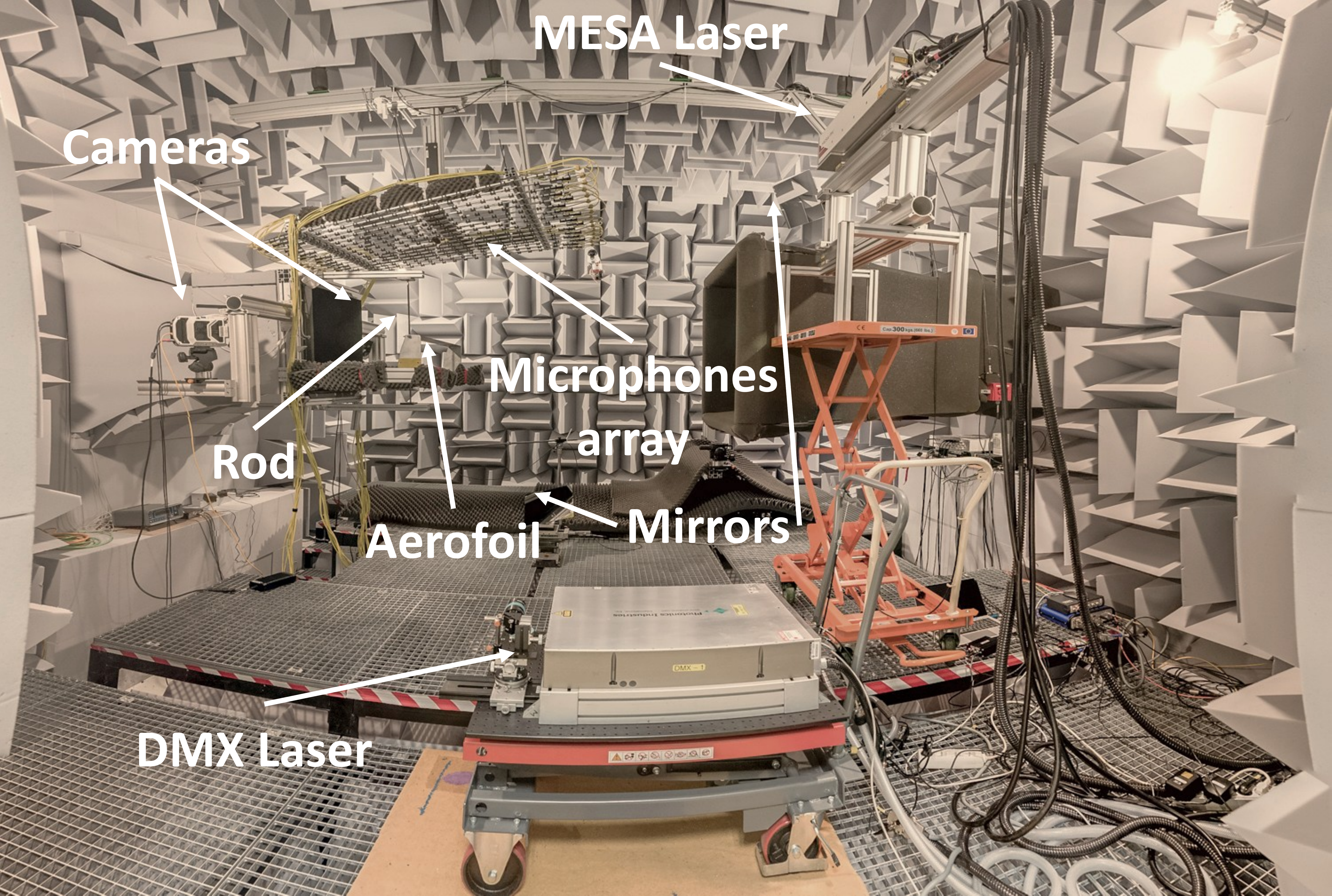}
	\caption{Picture of the experimental set-up for the $x_1$--$x_2$ (vertical) TR-PIV measurements.}
	\label{fig:photo_set-up}
\end{figure}

Illumination was recorded by two Vision Research Phantom v2640 cameras (1024 $\times$ 1024 pixels resolution), each coupled to a Nikon AF Micro-Nikkor 60 mm f/2.8D lens.
We used Scheimpflug mounts to ensure the laser sheet plane was entirely in focus.
An Antari Z-3000 aerosol generator was used to seed the air with heavy smoke (Algam Lighting FOG-LF-5L; a liquid formulation based on osmotic water and glycol 1,2-dihydroxypropane) -- the nominal size of the tracking particles is 2 $\mu$m.
Figure~\ref{fig:photo_set-up} shows a photograph of the experimental set-up at the BETI facility, including the lasers, cameras, microphone array, and other equipment used for the $x_1$--$x_2$ (vertical) measurements.
A similar set-up was used for the $x_1$--$x_3$ (horizontal) measurements, although the positioning of the cameras and lasers differed not shown here for brevity).

Table~\ref{tab:piv} presents the TR-PIV parameters, including the time between laser pulses ($t_\mathrm{exp}$), laser sheet thickness ($l_\mathrm{exp}$), acquisition frequency ($f_\mathrm{s}$), number of snapshots ($N_\mathrm{s}$), and approximate field of view (FoV) dimensions.
The table also lists the number of mesh points $N_x$ in the $x_1$--$x_2$ (vertical) and $x_1$--$x_3$ (horizontal) planes after processing the images using Davis software.
Figure~\ref{fig:sketch_PIV_acoustics} highlights the field of view for both planes in green.

\begin{table}[h]
	\caption{Parameters of the TR-PIV experiments. $^*x_d = x_2$ for the vertical plane and $x_d = x_3$ for the horizontal plane.}
	\begin{tabular}{c c c c c c c c c}
		\toprule
		Plane				& $t_\mathrm{exp}$ [$\mu$s]	& $l_\mathrm{exp}$ [mm]	& $f_\mathrm{s}$ [kHz]	& $N_\mathrm{s}$	& FoV ($L_\mathrm{x_d} \times L_\mathrm{x_1}$)$^*$	& $N_\mathrm{x_d} \times N_\mathrm{x_1}$$^*$	\\
		\midrule
		Vertical		& $40$											& $3$										& $6$										& $24,721$				& $3c \times 5c$																		& $136 \times 389$							\\
		Horizontal	& $40$											& $3$										& $6.25$								& $24,521$				& $4c \times 5c$																		& $266 \times 286$							\\
		\botrule
	\end{tabular}
	\label{tab:piv}
\end{table}

To process the raw images and obtain velocity fields, the average image over all acquisitions was first subtracted from each frame to enhance the signal-to-noise ratio by reducing background illumination.
A perspective correction was then applied to account for optical distortions (calibration + self-calibration procedures).
The initial interrogation pass used 64 $\times$ 64 pixel windows with 50\% overlap, while the final pass employed 16 $\times$ 16 pixel windows, also with 50\% overlap.
During the multi-pass post-processing, additional steps were taken, including the removal of vectors with correlation coefficients below 0.3 and the identification and replacement of outliers (universal outlier detection and replacement), along with other standard quality-control procedures.

\begin{figure}[h]
	\centering
	\includegraphics[width=0.8\textwidth]{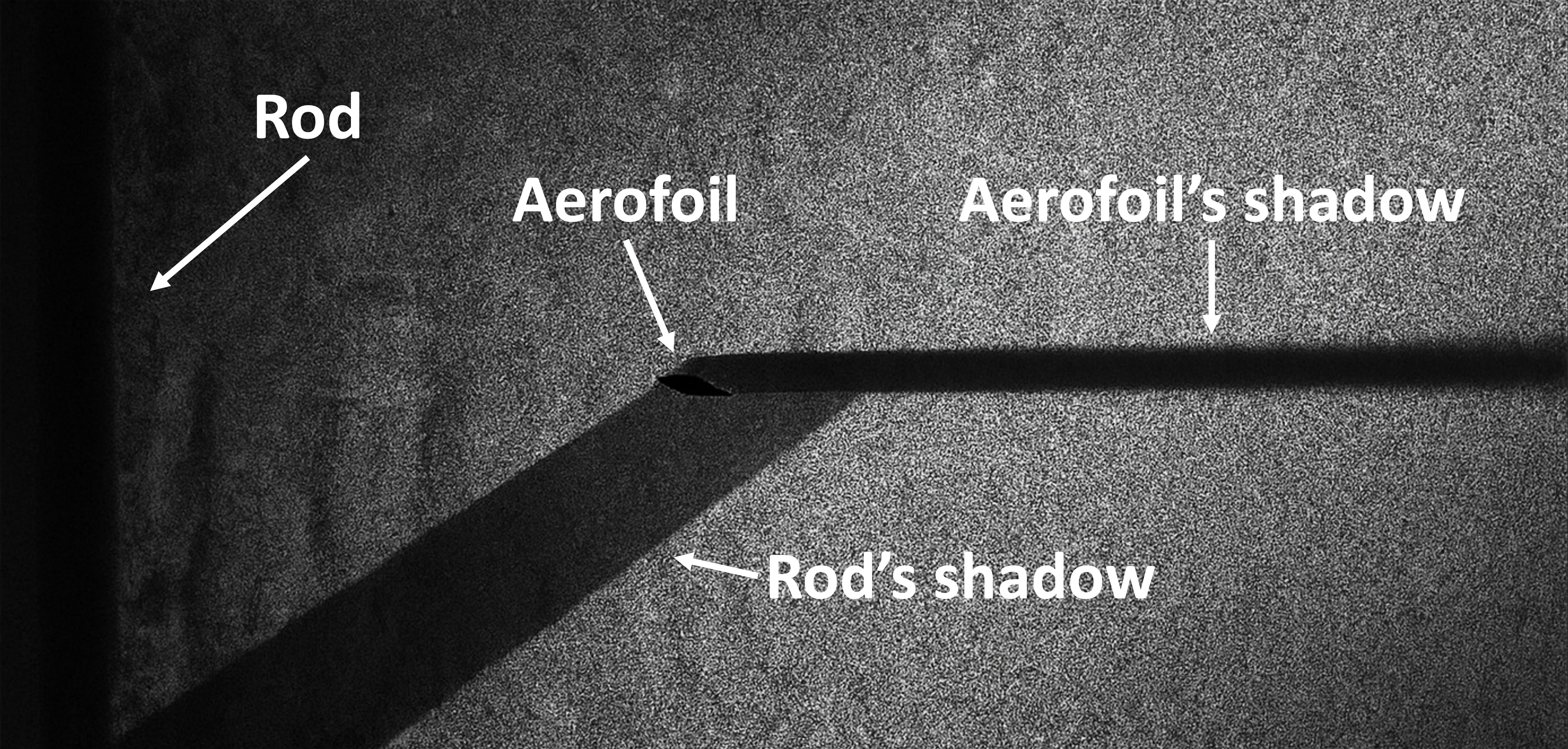}
	\caption{Raw image of a sample snapshot.}
	\label{fig:sample_snapshot_shadows}
\end{figure}

Figure~\ref{fig:sample_snapshot_shadows} shows a raw image of a random snapshot, prior to velocity field computation
One can observe shadows cast by the aerofoil and rod—regions that the camera cannot properly visualize due to the relative positioning of the cameras, laser, rod, and aerofoil.

Acoustic measurements were performed using an array of 256 ICS-43434 micro-electro-mechanical system (MEMS) microphones, each integrating a transducer, amplifier, and a 24-bit analog-to-digital converter.
These measurements were synchronized with the TR-PIV system via a trigger signal.
Further details on the MEMS array can be found in \citet{zhou2020design} and \citet{zhou2020three}.
In Figure~\ref{fig:sketch_PIV_acoustics}, the microphone array is depicted by black dots.
The acoustic signals were sampled at $f_\mathrm{s,a} = 50$~kHz and subsequently downsampled to match the TR-PIV acquisition frequency (see Table~\ref{tab:piv}).
Prior to downsampling, a Gaussian filter was applied to low-pass filter the signals.

%%%%%%%%%%%%%%%%%%%%%%%%%%%%%%%%%%%%%%%%%%%%%%
\subsection{Acoustic spectra}
\label{sec:methodology_acoustic}

The acoustic spectra for the microphones are computed using Welch's method~\citep{welch1967fft}.
The processing parameters include the block size ($N_\mathrm{fft}$), overlap block size ($N_\mathrm{ovlp}$), number of blocks ($N_\mathrm{b}$), Nyquist Strouhal number ($\mathrm{St}_\mathrm{Nyquist}$), and Strouhal number discretization ($\Delta\mathit{St}$), as listed in Table~\ref{tab:post_processing}, as well as a hamming window function.
The Strouhal number is a dimensionless parameter used to characterize oscillating flow phenomena, and it is defined as
\begin{equation}
	\mathit{St} = \frac{f d}{U_\infty} \mbox{,}
	\label{eq:strouhal}
\end{equation}
\noindent where $f$ is the frequency of oscillation, $d$ is the cylinder diameter, and $U_\infty$ is the free-stream speed.

\begin{table}[h]
	\caption{Welch post-processing parameters.}
	\begin{tabular}{c c c c c c }
		\toprule
		Plane				& $N_\mathrm{fft}$	& $N_\mathrm{ovlp}$	& $N_\mathrm{b}$	& $\mathit{St}_\mathrm{Nyquist}$	& $\Delta\mathit{St}$	\\
		\midrule
		Vertical		& $128$							& $64$							& $386$						& $2$															&	$0.031$							\\
		Horizontal	& $128$							& $64$							& $383$						& $2.083$													&	$0.033$							\\
		\botrule
	\end{tabular}
	\label{tab:post_processing}
\end{table}

The power spectral density (PSD), denoted by $\boldsymbol{\bar{P}}$ as a function of frequency, is computed as
\begin{equation}
	\boldsymbol{\bar{P}} = 10 \log \left[\frac{\left\langle\hat{p}^2\right\rangle}{{p_\mathrm{ref}}^2}\frac{U_\infty}{d}\right] \mbox{,}
	\label{eq:acoustic_spectra}
\end{equation}
\noindent where $\left\langle\hat{p}^2\right\rangle$ denotes the pressure power spectral density (PSD) estimate in the frequency domain, expressed in Pa$^2$/Hz.
The reference pressure is $p_\mathrm{ref} = 20~\mu$Pa.
The flow Mach number is defined as $M_\infty = U_\infty / a$, where $U_\infty$ is the free-stream velocity and $a = 343$~m/s is the ambient speed of sound.
The operator $\left\langle\cdot\right\rangle$ represents the expected value, or ensemble average.
This formulation yields a normalized spectral representation of the acoustic energy in dB/St.

%%%%%%%%%%%%%%%%%%%%%%%%%%%%%%%%%%%%%%%%%%%%%%
\subsection{Coherence}
\label{sec:methodology_coherence}

In order to provide insight into the interactions between aerodynamic structures and the generated noise, the linear link between the velocity and pressure fluctuations at any Strouhal number is investigated.
This is measured by the coherence between the pressure measured by a given microphone (e.g., the digital microphone above the aerofoil) and the velocity field at any mesh point $\boldsymbol{x} = \left(x_1, x_2, x_3\right)$ in the TR-PIV  grid, which is evaluated as
\begin{equation}
	\gamma_\mathit{p u_i}\left(\boldsymbol{x},\mathit{St}\right) = \frac{\left|S_\mathit{p u_i}\left(\boldsymbol{x},\mathit{St}\right)\right|}{\sqrt{S_\mathit{p p}\left(\mathit{St}\right) S_\mathit{u_i u_i}\left(\boldsymbol{x},\mathit{St}\right)}} \mbox{,}
	\label{eq:coherence}
\end{equation}
\noindent where $S_\mathrm{p u_i} = \left\langle \hat{p} \hat{u_i}^\dagger \right\rangle$ is the cross-spectral density between the acoustic pressure and any of the velocity components, $S_\mathit{p p} = \left\langle \hat{p}^2 \right\rangle$ is the power spectral density of the acoustic pressure signal, $S_\mathrm{u_i u_i} = \left\langle \hat{u_i}^2 \right\rangle$ (no summation implied) is the power spectral density of the velocity field, $\cdot^\dagger$ denotes the complex conjugate transpose and $\left\lvert\ \cdots \right\rvert$ denotes absolute values.
The coherence is evaluated separately for each velocity component $u_i$, with $u_1$, $u_2$, and $u_3$ representing the streamwise, upwash/downwash (cylinder span), and aerofoil span velocity components, respectively.

%%%%%%%%%%%%%%%%%%%%%%%%%%%%%%%%%%%%%%%%%%%%%%
\subsection{SPOD}
\label{sec:methodology_SPOD}

Coherent structures in the flow field, or in the acoustic pressure field, are identified using SPOD, which for every frequency provides the most energetic structures.
To perform SPOD, the first step is to move in the frequency domain~\citep{payne1967large, lumley1981coherent, bonnet1998collaborative, picard2000pressure} and compute the cross-spectral density (CSD) of the measurements, for every frequency of interest.
A state vector comprising both the velocity and pressure field is first defined as
\begin{equation}
	\boldsymbol{q} = \left[\boldsymbol{u}~p\right]^{T} \mbox{,}
	\label{eq:state_components}
\end{equation}
\noindent where $T$ denotes the transpose operator and $\boldsymbol{q} = \boldsymbol{q}\left(x_1,x_2,x_3,t\right)$.
It is assumed that the mean of each component has been subtracted so that $\boldsymbol{q}$ contains fluctuations.
After sampling in space and time (see Table~\ref{tab:piv}), $\boldsymbol{q}\in \mathbb{R}^{N_\mathrm{mesh} \times N_\mathrm{s}}$ is a matrix containing each of the $N_\mathrm{s}$ snapshots in its columns, each of size $N_\mathrm{mesh}$, where $N_\mathrm{mesh}$ is the sum of the TR-PIV grid points ($N_{x_d} \times N_{x_1} = 52,904$ and 76,076 for the vertical and horizontal planes, respectively) multiplied by three (due to the three velocity components) and the number of microphones ($N_\mathrm{mic} = 256$), i.e., $N_\mathrm{mesh} = 3 \times N_{x_d} \times N_{x_1} + N_\mathrm{mic} = 158,968$ (vertical plane) and 228,484 (horizontal plane).
%A Reynolds decomposition can be applied to the flow state components, i.e.,
%\begin{equation}
%	\boldsymbol{q} = \boldsymbol{\bar{q}} + \boldsymbol{q^{\prime}} \mbox{,}
%	\label{eq:reynolds_decomposition}
%\end{equation}
%\noindent where $\boldsymbol{\bar{q}}$ is the mean flow and $\boldsymbol{q^{\prime}}$ is the fluctuation.

The Fourier transform is applied to the state vector, that is, to the rows of $\boldsymbol{q}$.
Welch's method~\citep{welch1967fft} is employed for this task, as discussed in \S~\ref{sec:methodology_acoustic}.
We remark that tests with higher block sizes ($N_\mathrm{fft}$) were also performed, but the SPOD mode convergence was not satisfactory; these results are not shown here so as not to overburden the reader.
The choice of block size always involves a trade-off between frequency resolution and the spectral convergence of two-point statistics--and therefore of SPOD modes.
Accordingly, we chose to proceed with $N_\mathrm{fft} = 128$, following best practices of our research group when applying SPOD to TR-PIV data; see, for instance, \citet{cavalieri2013wavepackets}, \citet{jaunet2017two}, \citet{lesshafft2019resolvent}, and \citet{maia2024effect}, among others, who employed a similar TR-PIV setup for jet flow measurements.

For each frequency/Strouhal number, a matrix $\boldsymbol{\hat{q}} \in \mathbb{C}^{N_\mathrm{mesh} \times N_\mathrm{b}}$ is obtained~\citep{towne2018spectral}, a normalized version of which is
\begin{equation}
	\boldsymbol{\hat{Q}} = \frac{1}{\sqrt{N_\mathrm{b} \Delta\mathit{St}}} \boldsymbol{\hat{q}} \mbox{.}
	\label{eq:q_matrix}
\end{equation}
\noindent The matrix $\boldsymbol{\hat{Q}} \in\mathbb{C}^{N_\mathrm{mesh} \times N_\mathrm{b}}$ contains the state Fourier transforms of each data block normalized by spectral parameters $N_\mathrm{b}$ and $\Delta\mathit{St}$.
The multiplication of $\boldsymbol{\hat{Q}}$ with its adjoint performs the average over the blocks and provides the desired CSD, i.e.
\begin{equation}
	\boldsymbol{C} = \boldsymbol{\hat{Q}} \boldsymbol{\hat{Q}}^{\dagger} \mbox{.}
	\label{eq:csd}
\end{equation}

The classical SPOD problem now involves computing the eigenvalue decomposition of this CSD, weighted by an appropriate norm, frequency by frequency, i.e.,
\begin{equation}
	\boldsymbol{C} \boldsymbol{W} \boldsymbol{\Phi} = \boldsymbol{\Phi} \boldsymbol{\Lambda} \mbox{,}
	\label{eq:spod}
\end{equation}
\noindent where $\boldsymbol{W}$ is the norm, accounting for both the integration weights and the numerical quadrature on the discrete grid, $\boldsymbol{\Phi} \in \mathbb{C}^{N_\mathrm{mesh} \times N_\mathrm{mesh}}$ denotes the SPOD modes, and $\boldsymbol{\Lambda} \in \mathbb{R}^{N_\mathrm{mesh} \times N_\mathrm{mesh}}$ is a diagonal matrix with the positive eigenvalues $\lambda_i$ on its diagonal.
The SPOD modes form an orthonormal basis that optimally represents the CSD weighted by the chosen norm.
The contribution of each SPOD mode (eigenfunction) to the CSD is given by its corresponding eigenvalue, which ranks the modes by energetic content, i.e., from most to least energetic, with $\lambda_1 \geq \lambda_2 \geq \cdots \lambda_{n}$.

The size of matrix $\boldsymbol{C} \in \mathbb{C}^{N_\mathrm{mesh} \times N_\mathrm{mesh}}$ makes the eigenvalue decomposition in~\eqref{eq:spod} computationally expensive.
Therefore, instead of solving~\eqref{eq:spod}, we solve an equivalent problem that contains the same non-zero eigenvalues, given by
\begin{subeqnarray}
	\boldsymbol{\hat{Q}}^{\dagger} \boldsymbol{W} \boldsymbol{\hat{Q}} \boldsymbol{\Psi} &=&  \boldsymbol{\Psi} \boldsymbol{\Lambda} \mbox{,} \\
	\boldsymbol{\Phi} &=& \boldsymbol{\hat{Q}} \boldsymbol{\Psi} \boldsymbol{\Lambda}^{-1/2} \mbox{,}
	\label{eq:spod_snapshots}
\end{subeqnarray}
\noindent where $\boldsymbol{\hat{Q}}^{\dagger} \boldsymbol{W} \boldsymbol{\hat{Q}} \in \mathbb{C}^{N_\mathrm{b} \times N_\mathrm{b}}$, $\boldsymbol{\Psi} \in \mathbb{C}^{N_\mathrm{b} \times N_\mathrm{b}}$ denotes its eigenvectors, and $\boldsymbol{\Lambda} \in \mathbb{R}^{N_\mathrm{b} \times N_\mathrm{b}}$ contains the eigenvalues. 
$\boldsymbol{\Phi} \in \mathbb{C}^{N_\mathrm{mesh} \times N_\mathrm{b}}$ contains the SPOD modes in its columns, these are combinations of columns of $\boldsymbol{\hat{Q}}$ with expansion coefficients given by $\boldsymbol{\Psi}$.
The SPOD mode matrix can be written $\boldsymbol{\Phi} = [\boldsymbol{\Phi}_u~\boldsymbol{\Phi}_p]^T$, with velocity ($\boldsymbol{\Phi}_u$) and pressure ($\boldsymbol{\Phi}_p$) components.
The total number of non-zero eigenvalues is $N_\mathrm{b}$, which is the same as the number of blocks used in Welch's method to compute the CSDs, and this is the reason why a tilde ($\tilde{\cdot}$) is employed to indicate the SPOD modes and eigenvectors in this formulation and differentiate it from the classical approach.
This algorithm is known as the snapshots formulation of SPOD~\citep{sirovich1987turbulence, towne2018spectral, schmidt2020guide} and is computationally more efficient when $N_\mathrm{mesh} \gg N_\mathrm{b}$.

As noted by~\citet{freund2009turbulence}, the existence of SPOD modes does not necessarily imply dynamical significance.
It is essential to define an appropriate norm that effectively captures, for instance, the sound-producing dynamics of the flow.
Moreover, identifying a norm that yields a low-rank basis where only a few SPOD modes contain most of the flow energy can enable accurate reconstruction of flow statistics and capture the far-field acoustic signature.
Additionally, norms can be used to focus on specific spatial regions of the flow to maximize the energy associated with those regions~\citep{souza2019dynamics, kaplan2021nozzle}.

In this study, the database contains velocity and pressure measurements obtained in different spatial  domains, and we explore norms that maximize the energy of the structures in these domains.
For the velocity field, the turbulence kinetic energy (TKE) norm is given by~\citep{freund2009turbulence, schmidt2020guide}
\begin{equation}
	\boldsymbol{W}_{u} =
	\left[\begin{array}{cccc}
		\boldsymbol{K}_{u}	& \boldsymbol{Z}_{u}	& \boldsymbol{Z}_{u}	& \boldsymbol{Z}_{p}	\\
		\boldsymbol{Z}_{u}	& \boldsymbol{K}_{u}	& \boldsymbol{Z}_{u}	& \boldsymbol{Z}_{p}	\\
		\boldsymbol{Z}_{u}	& \boldsymbol{Z}_{u}	& \boldsymbol{K}_{u}	& \boldsymbol{Z}_{p}	\\
		\boldsymbol{Z}_{p}	& \boldsymbol{Z}_{p}	& \boldsymbol{Z}_{p}	& \boldsymbol{Z}_{p}	\\
	\end{array}\right] \mbox{,}
	\label{eq:spod_velocity_weights}
\end{equation}
\noindent where $\boldsymbol{K}$ is a diagonal matrix containing the quadrature weights and $\boldsymbol{Z}$ is the zero matrix.
Subscripts $u$ and $p$ denote velocity/TR-PIV and pressure/acoustic mesh grids, respectively.
The modes obtained using the TKE norm will be referred to as SPOD-u modes.
These modes maximize the kinetic energy in the flow field.
They contain a pressure component on the array which is merely subordinate to this flow field.
To focus on acoustic pressure, the operator $\boldsymbol{W}$ can be defined as~\citep{freund2009turbulence, amaral2024pressure}
\begin{equation}
	\boldsymbol{W}_{p} =
	\left[\begin{array}{cccc}
		\boldsymbol{Z}_{u}	& \boldsymbol{Z}_{u}	& \boldsymbol{Z}_{u}	& \boldsymbol{Z}_{p}	\\
		\boldsymbol{Z}_{u}	& \boldsymbol{Z}_{u}	& \boldsymbol{Z}_{u}	& \boldsymbol{Z}_{p}	\\
		\boldsymbol{Z}_{u}	& \boldsymbol{Z}_{u}	& \boldsymbol{Z}_{u}	& \boldsymbol{Z}_{p}	\\
		\boldsymbol{Z}_{p}	& \boldsymbol{Z}_{p}	& \boldsymbol{Z}_{p}	& \boldsymbol{K}_{p}	\\
	\end{array}\right] \mbox{.}
	\label{eq:spod_pressure_weights}
\end{equation}
The modes obtained using the acoustic pressure norm will be referred to as SPOD-p modes.
These modes maximize the pressure energy at the microphones array.
%They contain a velocity component in the flow field, which is subordinate to their acoustic pressure component.
In the following, SPOD-p modes are used to infer the velocity structures connected to the most energetic acoustic modes.

Appendix~\ref{app:methodology_ESPOD} presents the equivalence between SPOD weighted by a proper norm $\boldsymbol{W}$ and the extended SPOD (ESPOD)~\citep{boree2003extended}.

Regarding modes convergence, following~\citet{cavalieri2013wavepackets, lesshafft2019resolvent}, the alignment between an SPOD mode evaluated using the full data series ($\boldsymbol{\Phi}_\mathrm{full}$) and the first and second halves ($\boldsymbol{\Phi}_i$, with $i = 1$ and 2 for the first and second halves of the time series, respectively), is given as
\begin{equation}
    \beta(\mathit{St}) = \frac{\left\lvert\ \langle \boldsymbol{\Phi}_\mathrm{full}(\mathit{St},\boldsymbol{x}),~{\boldsymbol{\Phi}_i}(\mathit{St},\boldsymbol{x}) \rangle \right\rvert}{\vert\vert \boldsymbol{\Phi}_\mathrm{full}(\mathit{St},\boldsymbol{x})\vert\vert \hspace{2pt} \vert\vert {\boldsymbol{\Phi}_i}(\mathit{St},\boldsymbol{x})\vert\vert} \mbox{,}
\label{eq:beta}
\end{equation}
\noindent where $\langle \cdot,\cdot \rangle$ denotes the inner product between SPOD modes computed using the full and half time series and $||\cdot||$ denotes the Euclidean norm.
Perfectly aligned, or converged, modes have $\beta = 1$, whereas a complete misalignment is given by $\beta = 0$.

%%%%%%%%%%%%%%%%%%%%%%%%%%%%%%%%%%%%%%%%%%%%%%
\subsection{Conventional beamforming}
\label{sec:conventional_beamforming}

Beamforming algorithms enable the mapping of acoustic source locations and power levels based on measurements obtained from a microphone array~\citep{mueller2002aeroacoustic, chiariotti2019acoustic, merinomartinez2019review}.
The conventional beamforming formulation used in the present study, in the frequency domain, is given by
\begin{equation}
	b(\boldsymbol{r}_{o,n},\mathit{St}) = \boldsymbol{h}^\dagger(\boldsymbol{r}_{m,n},\mathit{St}) \left\langle \hat{\boldsymbol{p}}(\boldsymbol{r}_{m^\prime,o},\mathit{St}) \hat{\boldsymbol{p}}^\dagger(\boldsymbol{r}_{m,o},\mathit{St}) \right\rangle \boldsymbol{h}(\boldsymbol{r}_{m,n},\mathit{St}) \mbox{,}
	\label{eq:beamforming}
\end{equation}
\noindent where $b$ is the source power level at a given position $\boldsymbol{r}_{o,n}$, with $\boldsymbol{r}_{o,n}$ denoting the vector between the microphone array barycentre/centroid $o$ and the scanning mesh point $n$ ($1 \leq n \leq N$).
The vector $\boldsymbol{h}$ is the steering vector for a given position $\boldsymbol{r}_{m,n}$, with $m$ ($1 \leq m \leq M$) denoting a microphone in the array, and $\hat{\boldsymbol{p}}$ indicates the measured pressure in the frequency domain at a microphone located at $\boldsymbol{r}_{m,o}$.
Beamforming is performed for each frequency $\mathit{St}$.
Recall that $\mathit{St}$ is defined in~\eqref{eq:strouhal}, and the angular frequency is given by $\omega = 2\pi f$.

The term
\begin{equation}
	\boldsymbol{P} = \left\langle \hat{\boldsymbol{p}}(\boldsymbol{r}_{m,o},\mathit{St}) \hat{\boldsymbol{p}}^\dagger(\boldsymbol{r}_{m^\prime,o},\mathit{St}) \right\rangle \mbox{,}
	\label{eq:CSM}
\end{equation} 
\noindent denotes the cross-spectral matrix (CSM), or CSD, of microphone signals.
The CSM is computed using Welch's method, with parameters summarized in Table~\ref{tab:post_processing}, and the frequencies are then grouped into one-third octave bands.

Following~\citet{padois2013numerical}, a convective correction is applied to the streamwise direction of the transfer function to account for the flow in open-section tunnels, i.e.,
\begin{equation}
	\boldsymbol{r}_{m,n} = ({x_1}_{m,n} + M_\infty H) \hat{i} + {x_2}_{m,n} \hat{j} + {x_3}_{m,n} \hat{k} \mbox{,}
	\label{eq:convective_correction}
\end{equation}
\noindent where $(\hat{i},\hat{j},\hat{k})$ denote the unit vectors in the three Cartesian directions, $({x_1}_{m,n}, {x_2}_{m,n}, {x_3}_{m,n})$ are the three components of the vector between a microphone $m$ and a scanning mesh point $n$, and $H$ is the distance between the scanning mesh and the tunnel mixing layer (tunnel nozzle half-height).

The steering vector can be interpreted as a normalized Green's function, and the following definition is employed~\citep{pagani2016slat, pagani2019experimental}
\begin{equation}
	\boldsymbol{h}(\boldsymbol{r}_{m,n},\mathit{St}) = \frac{\boldsymbol{g}(\boldsymbol{r}_{m,n},\mathit{St})}{\sqrt{\sum_{m=1}^M \sum_{m^\prime=1}^M |\boldsymbol{g}(\boldsymbol{r}_{m,n},\mathit{St})| |\boldsymbol{g}(\boldsymbol{r}_{m^\prime,n},\mathit{St})|}} \mbox{,}
	\label{eq:steering_vector}
\end{equation}
\noindent where $\boldsymbol{g}(\boldsymbol{r}_{m,n},\omega)$ is the monopole Green's function normalized to provide values at the microphone array barycentre/centroid $o$, i.e.,
\begin{equation}
	\boldsymbol{g}(\boldsymbol{r}_{m,n},\mathit{St}) = \frac{{r}_{o,n}}{{r}_{m,n}} e^{\frac{-i\omega({r}_{m,n}-{r}_{o,n})}{a}} \mbox{,}
	\label{eq:green_function}
\end{equation}
\noindent where non-bold notation for ${r}_{o,n}$ and ${r}_{m,n}$ denotes absolute (scalar) values.
It is important to emphasize that other steering vector formulations can be adopted, slightly modifying the sources location and/or power~\citep{sarradj2012three}.

Beamforming results are expressed using the same normalization as in~\eqref{eq:acoustic_spectra}, as follows
\begin{equation}
	B(\boldsymbol{r}_{o,n},\mathit{St}) = 10 \log \left[\frac{b(\boldsymbol{r}_{o,n},\mathit{St})}{{p_\mathrm{ref}}^2}\frac{U_\infty}{d}\right] \mbox{.}
	\label{eq:beamforming_dBSt}
\end{equation}

\begin{figure}[h]
	\centering
		\includegraphics[width=0.65\textwidth]{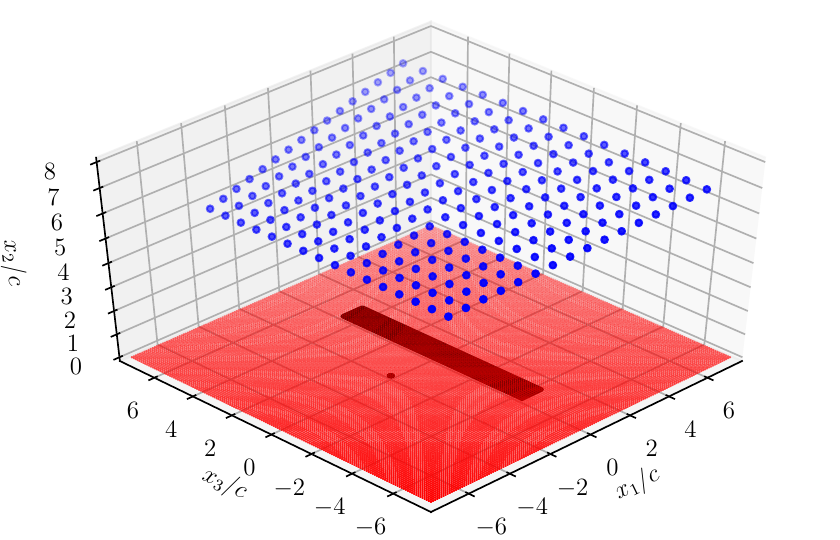}
	\caption{Sketch of the beamforming scanning mesh (orange dots), the microphone array (blue closed circles), the perpendicular rod (black closed circle), and the aerofoil (continuous line rectangle) projections on the mesh.}
	\label{fig:beamforming_sketch}
\end{figure}

In the present study, the beamforming scanning mesh is defined in the aerofoil chord-span plane ($x_1$--$x_2$), and it is centred at the aerofoil leading-edge mid-span.
The mesh spans $15c$ (1500~mm) in both the $x_1$ and $x_2$ directions, with a discretization of $0.1c$ (10~mm) in both directions.
Figure~\ref{fig:beamforming_sketch} presents a sketch of the beamforming domain employed in this study. 

Appendix~\ref{app:bw_dr} presents the characteristics of the microphone array, such as resolution and dynamic range as a function of frequency.

A low-rank model of the microphone array CSM ($\boldsymbol{\Pi}{k}$) can be obtained using the pressure component of the SPOD modes as
\begin{equation}
	\boldsymbol{\Pi}_{k} = \boldsymbol{\Phi}_{k} \boldsymbol{\Lambda}_{k} {\boldsymbol{\Phi}_{k}}^\dagger \mbox{,}
	\label{eq:CSM_low-rank}
\end{equation}
\noindent where the subscript $k$ indicates the subset of modes retained to obtain the low-rank CSM, e.g., the rank-1 or 2 SPOD mode, or a number of SPOD modes.
$\boldsymbol{\Phi}$ and $\boldsymbol{\Lambda}$ are the SPOD modes and corresponding eigenvalues, respectively, as defined in~\eqref{eq:spod_snapshots}.

It is thus possible to evaluate beamforming informed by a low-rank model of the microphone array CSM as
\begin{equation}
	b(\boldsymbol{r}_{o,n},\mathit{St}) = \boldsymbol{h}^\dagger(\boldsymbol{r}_{m,n},\mathit{St}) \boldsymbol{\Pi}_{k} \boldsymbol{h}(\boldsymbol{r}_{m,n},\mathit{St}) \mbox{.}
	\label{eq:beamforming_low-rank}
\end{equation}

%%%%%%%%%%%%%%%%%%%%%%%%%%%%%%%%%%%%%%%%%%%%%%
\section{Results}
\label{sec:results}
%%%%%%%%%%%%%%%%%%%%%%%%%%%%%%%%%%%%%%%%%%%%%%

%%%%%%%%%%%%%%%%%%%%%%%%%%%%%%%%%%%%%%%%%%%%%%
\subsection{Acoustic spectra}
\label{sec:results_acoustic}

\begin{figure}[h]
	\centering
	\begin{subfigure}{0.49\textwidth}
		\includegraphics[width=\textwidth]{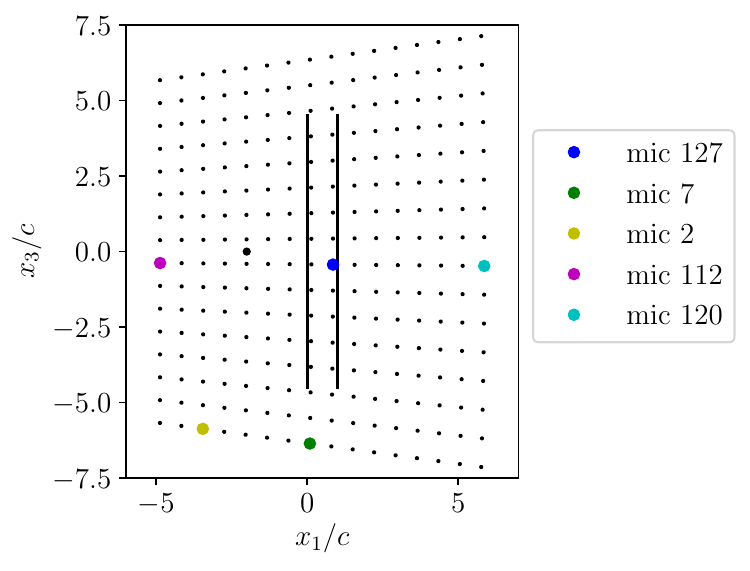}
		\caption{Experimental set-up top view}
		\label{fig:acoustic_microphones}
	\end{subfigure}
	\begin{subfigure}{0.49\textwidth}
		\includegraphics[width=\textwidth]{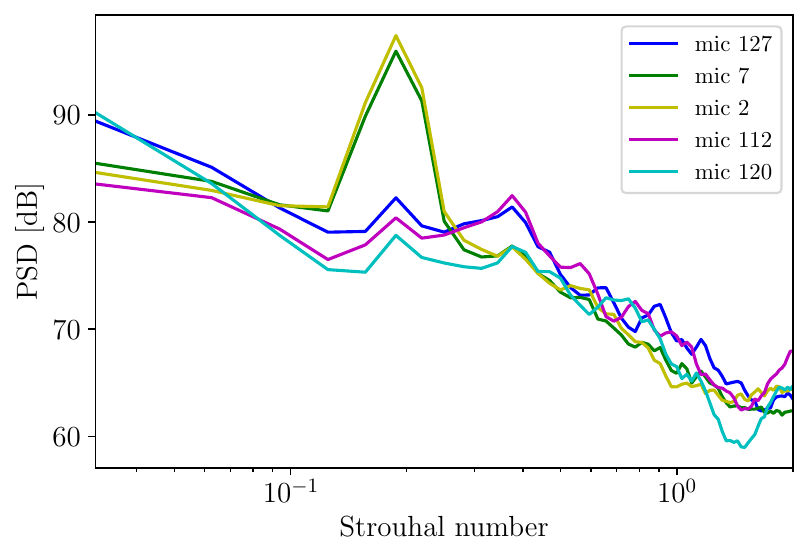}
		\caption{Spectra}
		\label{fig:acoustic_spectra}
	\end{subfigure}
	\caption{Acoustic spectra for selected microphones. Left frame: digital microphones array (small dots), cylinder (closed black circle), aerofoil (vertical black lines), and the selected microphones (coloured markers). Right frame: acoustic spectra.}
	\label{fig:acoustic}
\end{figure}

Figure~\ref{fig:acoustic} presents the acoustic spectra for selected microphones in the digital array.
The left frame of the figure shows a top view of the experimental set-up, with coloured markers indicating the microphones selected for spectral analysis.
Microphones located approximately at the midspan position of the aerofoil were chosen: one microphone positioned upstream of the rod/cylinder (magenta colour, mic \#112), another over the aerofoil (blue colour, mic \#127),  and a third downstream of the aerofoil (cyan colour, mic \#120).
Additionally, spectra from two other microphones -- one located at the side of the cylinder (yellow, colour mic \#2) and the other near the aerofoil side (green colour, mic \#7) -- are also included.

The microphones aligned with the aerofoil midspan (mic \#112, \#127, and \#120 -- magenta, blue, and cyan colours, respectively) exhibit spectra with similar shapes and intensity levels.
However, the microphone located downstream of the aerofoil's trailing edge (mic \#120, cyan colour) shows lower intensity levels than the other two, especially in the range $0.1 \lesssim \mathit{St} \lesssim 0.6$.
The von Kármán vortex shedding frequency ($\mathit{St} \approx 0.2$) is more pronounced for the microphone positioned over the aerofoil (mic \#127, blue colour), while its harmonics ($\mathit{St} \approx 0.4$) have a more substantial signature on the microphone located upstream of the cylinder (mic \#112, magenta colour).
Note that, for $N_b = 386$, the uncertainty in the spectral levels is approximately 0.4 dB/St at a 95\% confidence interval.
Such a value was calculated following the procedures described in \citet{bendat2011random}, where the lower and upper bounds in dB are given by
\begin{subeqnarray}
	P_\mathit{lower} &=& 10\log_{10}\left(\frac{2 N_b}{\chi^2_{2 N_b,0.025}}\right) \mbox{,} \\
	P_\mathit{upper} &=& 10\log_{10}\left(\frac{2 N_b}{\chi^2_{2 N_b,0.975}}\right) \mbox{,}
	\label{eq:uncertainty}
\end{subeqnarray}
\noindent where, for $N_b = 386$, $P_\mathit{lower} = -0.42$ dB ($\chi^2_{2 N_b,0.025} = 696.897$) and $P_\mathit{upper} = 0.44$ dB ($\chi^2_{2 N_b,0.975} = 850.891$).

The two lateral microphones, positioned on the cylinder and aerofoil sides (mics \#2 and \#7 -- yellow and green colours, respectively), exhibit very similar spectra in terms of shape and intensity levels.
The peak at $\mathit{St} \approx 0.4$ is less evident for these microphones than for the midspan microphones.
%These results align with previous tandem cylinder literature~\citep{eltaweel2011numerical, giret2015noise}, which has shown that at $\mathit{St} \approx 0.2$, the dipole directivity points toward 90 degrees relative to the aerofoil chord.
%Conversely, at $\mathit{St} \approx 0.4$, the strongest levels of the dipolar pattern are observed perpendicular to the aerofoil, aligned with the aerofoil chord, as will be shown in \S~\ref{sec:results_beamforming}.

%%%%%%%%%%%%%%%%%%%%%%%%%%%%%%%%%%%%%%%%%%%%%%
\subsection{Coherence between acoustic and velocity fields}
\label{sec:results_coherence}

Coherence is now exploited to find out if some velocity components in the flow field are connected to the observed acoustic peaks.

\begin{figure}[h]
	\centering
	\includegraphics[width=\textwidth]{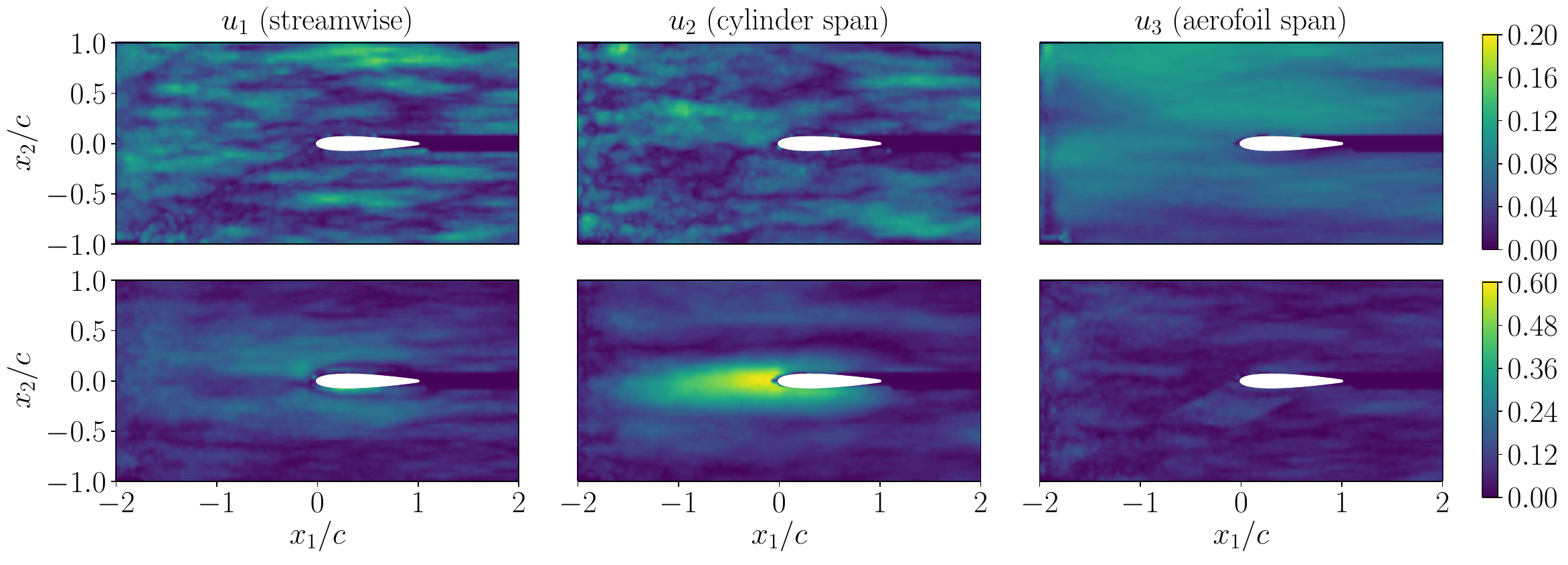}
	\caption{Coherence between microphone \#127 and velocity components for the vertical plane. The aerofoil is represented by the white shaded area, while the rod trailing edge is located at $x_1/c = -2$ spanning the entire $x_2$ direction. Frames, from left to right: streamwise ($u_1$), upwash/downwash ($u_2$) and aerofoil span ($u_3$) velocity components, respectively. Top frames denote $\mathit{St} \approx 0.2$ and bottom frames denote $\mathit{St} \approx 0.4$.}
	\label{fig:coherence_vertical}
\end{figure}

\begin{figure}[h]
	\centering
	\includegraphics[width=\textwidth]{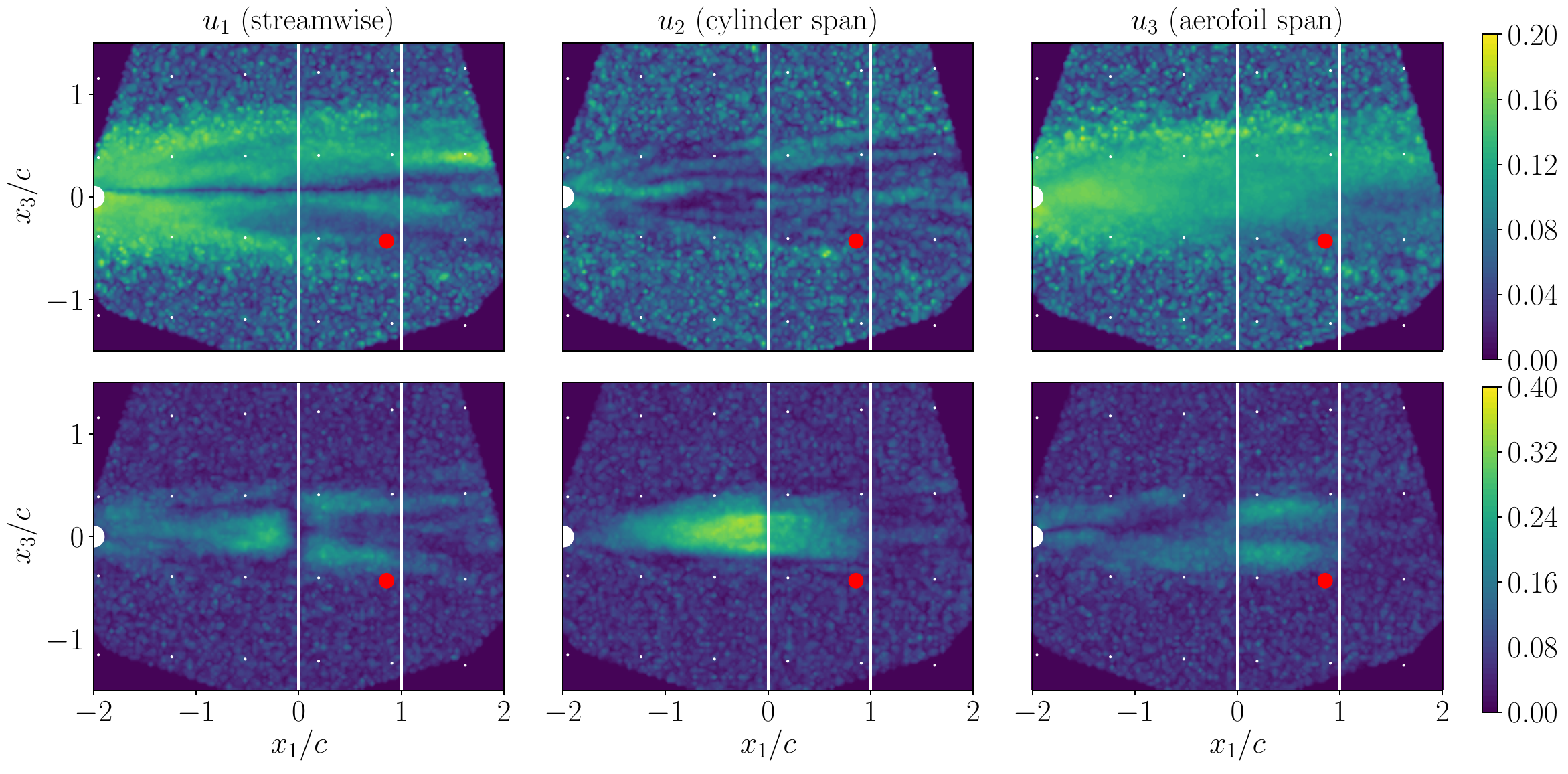}
	\caption{Coherence between the microphone \#127 and velocity components for the horizontal plane. The closed circle and the two vertical lines indicate the positions of the rod and the aerofoil, respectively, whereas the dots denote microphones. Microphone \#127 is highleted in red. See the comments in the caption of Figure~\ref{fig:coherence_vertical}.}
	\label{fig:coherence_horizontal}
\end{figure}

Figures~\ref{fig:coherence_vertical} and \ref{fig:coherence_horizontal} show the coherence levels~\eqref{eq:coherence} for the vertical and horizontal planes (Figure~\ref{fig:sketch_PIV_acoustics}), respectively.
Results for $\mathit{St} \approx 0.2$ and 0.4 are highlighted in the figures, as these frequencies coincide with peaks observed in the noise spectra (Figure~\ref{fig:acoustic_spectra}).
The vertical plane (Figure~\ref{fig:coherence_vertical}) displays strong coherence values, reaching up to 60\% between microphone \#127 (located approximately above the mid-chord, mid-span position of the aerofoil, so to capture the effects of wake/aerofoil interaction) and the $u_2$ velocity component at $\mathit{St} \approx 0.4$, particularly near the aerofoil leading edge.

This finding is consistent with prior literature that identifies the upwash/downwash velocity component~\citep{roger2014vortex, quaglia20173d, zehner2018aeroacoustic} and the region near the aerofoil leading edge~\citep{jacob2005rod, boudet2005wake} as responsible for the rod/aerofoil interaction noise.
For $\mathit{St} \approx 0.2$, maximum coherence levels of approximately 20\%, with no clear dominance of any single velocity component; however, the $u_3$ component shows more distributed coherence in the upper part of the domain, with values as high as 10\%.

Regarding the horizontal plane coherence plots (Figure~\ref{fig:coherence_horizontal}), the $u_1$ and $u_3$ velocity components, associated with the von Kármán vortex shedding, exhibit the highest coherence levels -- up to 20\% -- at $\mathit{St} \approx 0.2$.
Conversely, the $u_2$ velocity component shows the highest coherence levels, reaching up to 40\% at $\mathit{St} \approx 0.4$, which is consistent with the results in the vertical plane.

Overall, the $u_2$ component exhibits the highest coherence levels at $\mathit{St} \approx 0.4$ (double the von Kármán vortex shedding frequency) for both planes, reaching 60\% in the vertical plane and 40\% in the horizontal plane.
This indicates that the frequency $\mathit{St} \approx 0.4$ is associated with rod wake/aerofoil interaction noise.
In contrast, for $\mathit{St} \approx 0.2$, higher coherence levels of up to 20\% are observed in the horizontal plane, particularly for the $u_1$ and $u_3$ velocity components, both of which are linked to the rod wake.

%%%%%%%%%%%%%%%%%%%%%%%%%%%%%%%%%%%%%%%%%%%%%%
\subsection{SPOD analysis}
\label{sec:results_structures}

This section outlines the data post-processing procedures applied to the TR-PIV and microphone array measurements. TR-PIV provides access to the near-field turbulent structures.
The application of SPOD-u enables the identification of coherent physical structures embedded in the turbulent flow, such as the von Kármán vortex street.
Additionally, SPOD-p is used to correlate the near-field turbulence with the acoustic field, thereby isolating flow structures that are acoustically relevant and contribute to sound generation.

%%%%%%%%%%%%%%%%%%%%%%%%%%%%%%%%%%%%%%%%%%%%%%
\subsubsection{SPOD convergence}
\label{sec:results_convergence}

\begin{figure}
	\centering
	\includegraphics[width=\textwidth]{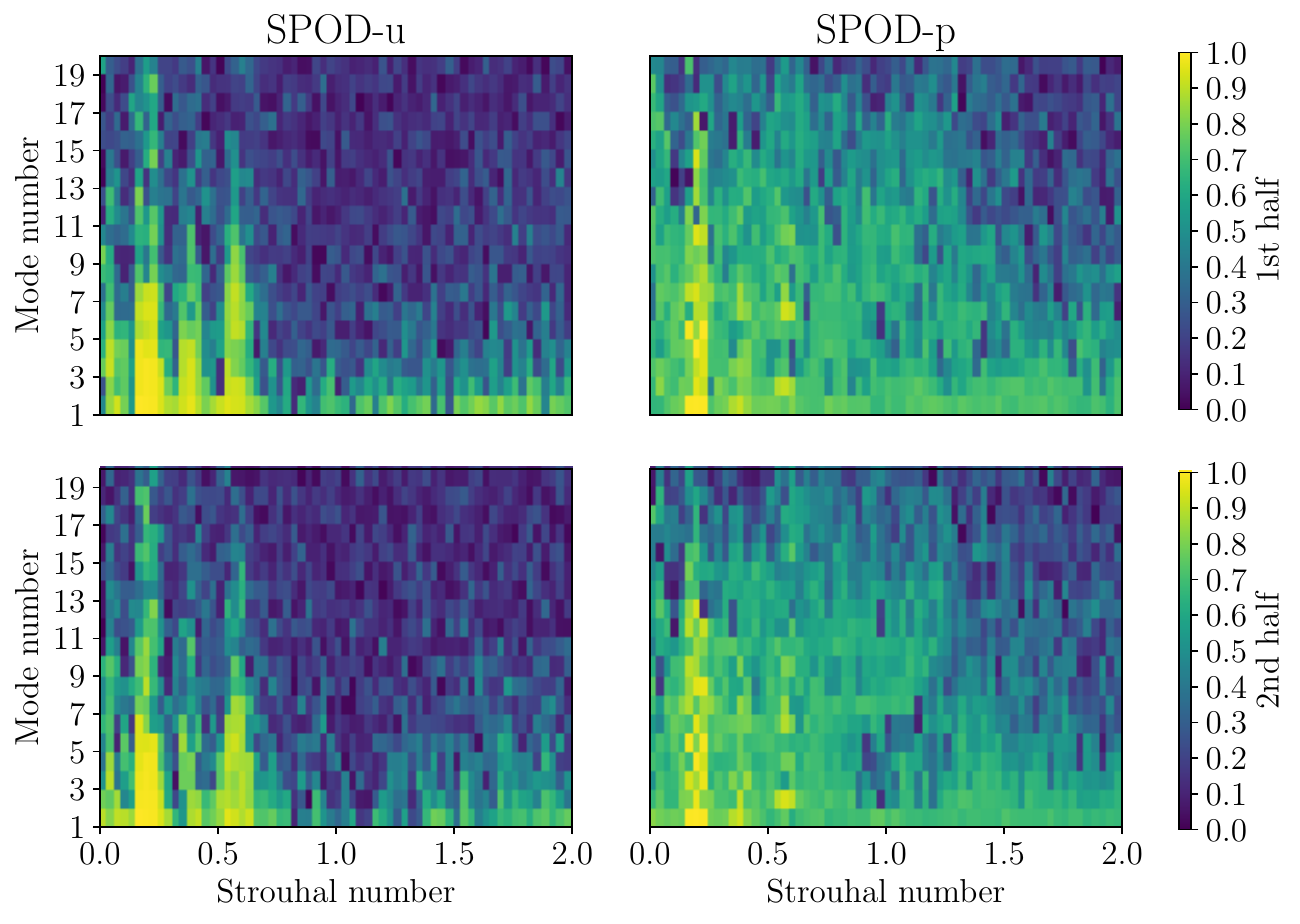}
	\caption{SPOD convergence levels for the vertical plane configuration ($x_1$--$x_2$). Contour maps of convergence levels $\beta$ computed using the first (top panels) and second (bottom panels) halves of the time series, for SPOD-u (left) and SPOD-p (right).}
	\label{fig:spod_convergence_vertical}
\end{figure}

Figure~\ref{fig:spod_convergence_vertical} shows the convergence levels~\eqref{eq:beta} of the SPOD calculations for the first 20 modes, using both the pressure (SPOD-p) and the TKE (SPOD-u) norms for the vertical configuration ($x_1$--$x_2$ plane).
The first few modes appear to be well-converged for both norms, with $\beta$-values very close to 1, especially at $\mathit{St} \approx 0.2$, 0.4, and 0.6.
However, the SPOD-p modes exhibit convergence over a broader region in the mode--Strouhal number space when compared to SPOD-u.

\begin{figure}
	\centering
	\includegraphics[width=\textwidth]{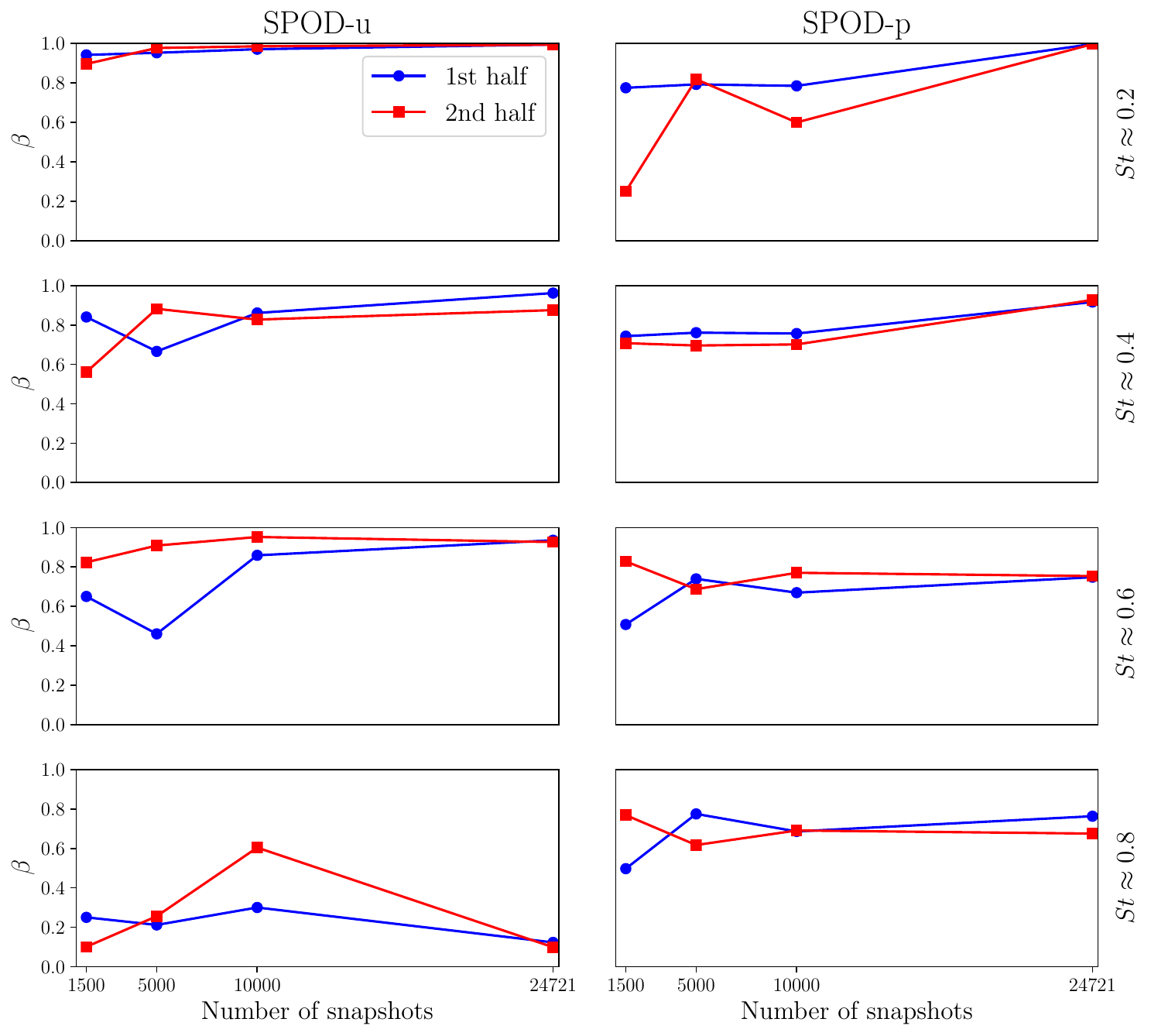}
	\caption{SPOD convergence levels for the vertical plane configuration ($x_1$--$x_2$). Left and right panels show SPOD-u and SPOD-p, respectively, while rows from top to bottom correspond to $\mathit{St} \approx 0.2$, 0.4, 0.6, and 0.8. Blue curves with circular markers indicate convergence levels $\beta$ computed using the first half of the time series, whereas red curves with square markers correspond to the second half.}
	\label{fig:spod_convergence2_vertical}
\end{figure}

A further step is to reduce the size of the time series and verify the corresponding convergence levels.
This is shown in Figure~\ref{fig:spod_convergence2_vertical}, where the convergence levels are plotted for selected frequencies using the full number of snapshots (24,721), as well as subsets of 10,000, 5,000, and 1,500 snapshots.
In practice, this reduction is equivalent to decreasing the number of blocks $N_b$ employed to compute the SPOD.
Overall, it is observed that, particularly for SPOD-p and at lower frequencies ($\mathit{St} \approx 0.2$ and 0.4), convergence levels degrade significantly even with 10,000 snapshots, i.e. using roughly 1.5 times fewer blocks than in the full case.
In this situation, convergence drops from values in the 0.95--1 range to values in the 0.6--0.8 range.
For SPOD-u, the loss of coherence becomes more evident only when the number of snapshots is further reduced, e.g. to 5,000.

Convergence of two-point statistics—and consequently of SPOD modes—is always a challenging task.
The algorithm relies on Welch’s method, which processes a finite time series of length $N_s$.
By selecting Welch parameters such as $N_\mathit{fft}$ (block size) and block overlap, one must balance frequency resolution against spectral convergence.
Increasing the number of blocks ($N_b$) improves the convergence of two-point statistics but reduces frequency resolution, and the reverse also holds.
The reader is referred to \citet{blanco2022improved} and \citet{heidt2024optimal} for a detailed discussion of SPOD mode convergence, which lies beyond the scope of the present manuscript.
Typically, only the first few modes can be considered well converged, while the less energetic modes are less reliable--a trend also reported in jet flow databases (see, for example, \citet{lesshafft2019resolvent} and \citet{amaral2025tabbed}).
In the present case, however, at $\mathit{St} \approx 0.2$ and 0.4—the two main frequencies of interest—the first five modes exhibit good convergence.
Since our focus is primarily on the first two modes to elucidate the physical mechanisms underlying wake/aerofoil interaction noise, the achieved convergence levels are deemed satisfactory.

%%%%%%%%%%%%%%%%%%%%%%%%%%%%%%%%%%%%%%%%%%%%%%
\subsubsection{SPOD eigenvalues}
\label{sec:results_eigenvalues}

\begin{figure}[h]
	\centering
	\includegraphics[width=\textwidth]{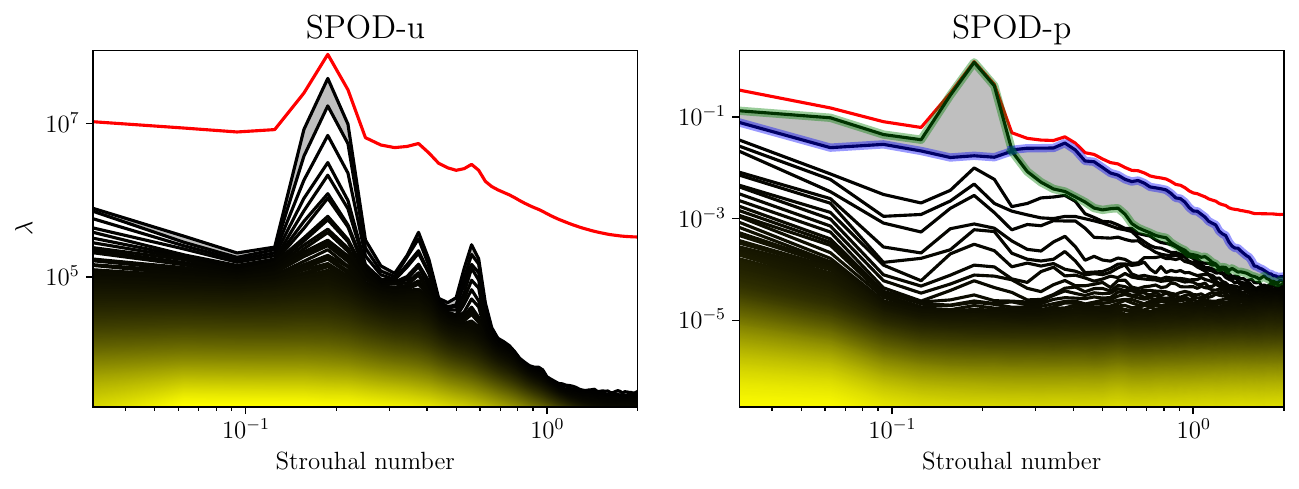}
	\caption{SPOD energy spectra for the vertical plane configuration. The red colour represents the total energy, while the colour gradient from black to yellow denotes the spectrum of each rank, from the most to the least energetic. The grey area indicates the separation between the rank-1 and 2 modes. The blue curve on the right plane denotes the spectra related to the aerofoil structures, whereas the green curve indicates the structures related to the rod.}
	\label{fig:spectra_vertical}
\end{figure}

Let us begin by analysing the structures extracted from the TR-PIV vertical plane, i.e. perpendicular to the aerofoil.
Figure~\ref{fig:spectra_vertical} shows the SPOD eigenvalues spectra for the TKE and pressure norms, i.e. SPOD-u and SPOD-p, respectively.
In this figure, the red colour represents the total energy spectra, while the colour gradient from black to yellow denotes the modes from the most to the least energetic.
The closer the total energy curve (red curve) is to the rank-1 mode (top black curve), the more low-rank the spectra are.
The shaded grey area between the rank-1 and 2 modes highlights the dominance of the rank-1 mode over the rank-2.

Spectral peaks at $\mathit{St} \approx 0.2$, 0.4, and 0.6 are clearly visible in the SPOD-u spectra (left subplot) for the first few more energetic modes.
Note that $\mathit{St} \approx 0.2$ is connected to the von Kármán vortex shedding frequency, and $\mathit{St} \approx 0.4$ corresponds to the rod wake/aerofoil interaction, as addressed in \S~\ref{sec:results_coherence}.
The SPOD-u spectra are not low-rank~\citep{schmidt2020guide}, except around $\mathit{St} \approx 0.2$, where the rank-1 mode accounts for approximately 50\% of the total energy.

The results differ when considering the SPOD-p spectra.
Clear peaks are observed only at $\mathit{St} \approx 0.2$ and 0.4, and both frequencies exhibit low-rank behaviour, with the rank-1 mode accounting for approximately 95\% and 75\% of the flow energy, respectively.
The blue curve shown in the SPOD-p spectra follows the rank-2 mode up to $\mathit{St} \approx 0.3$ and the rank-1 mode beyond this Strouhal number.
It appears that the structures identified with the rank-1 SPOD mode for $\mathit{St} \gtrsim 0.3$ are actually tracked in the rank-2 SPOD mode for $\mathit{St} \lesssim 0.3$.
Upon closer inspection of the spectra, it can be observed that for $\mathit{St} \lesssim 0.3$, the rank-2 and 3 SPOD modes also exhibit a significant energy separation.
These structures are hypothesized to be connected to the perpendicular-rod–aerofoil impingement noise, as will be discussed later, and will be refereed as aerofoil branch.
Conversely, the structures contained in the rank-1 SPOD mode for $\mathit{St} \lesssim 0.3$ (green curve) and in the rank-2 mode for $\mathit{St} \gtrsim 0.3$ are associated with the cylinder lift fluctuation (isolated-cylinder noise) and will be referred to as the cylinder branch.
This kind of behaviour is also observed in jets within the resolvent framework (response and forcing modes), with different mechanisms, such as Kelvin--Helmholtz (K--H) and Orr, being tracked in suboptimal modes depending on the Strouhal range~\citep{schmidt2018spectral, lesshafft2019resolvent}.

\begin{figure}
	\centering
	\includegraphics[width=\textwidth]{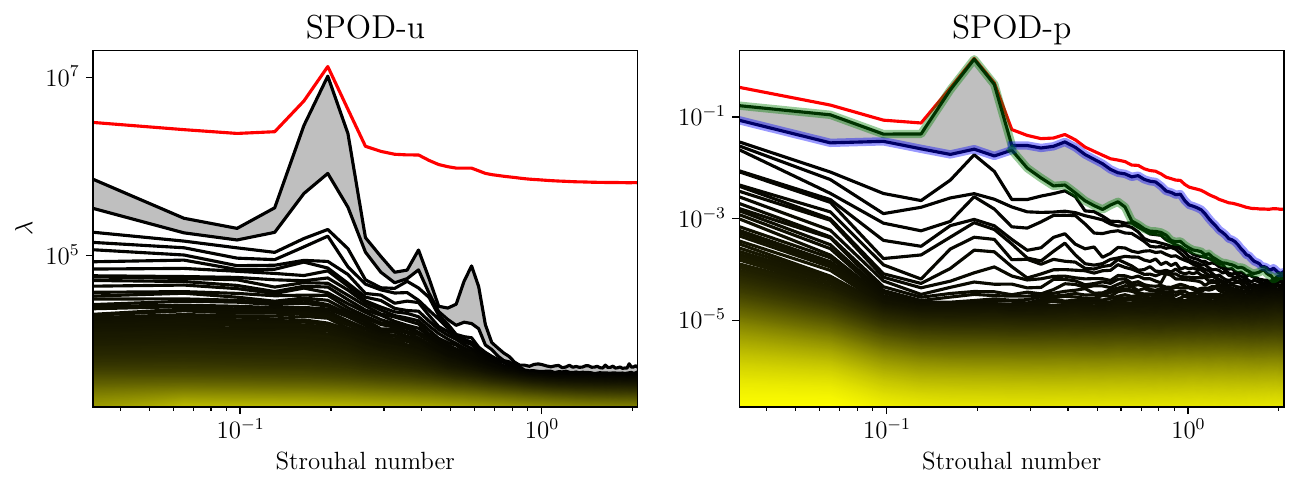}
	\caption{SPOD and energy spectra for the horizontal plane configuration ($x_1$--$x_3$). See the comments in the caption of Figure~\ref{fig:spectra_vertical}.}
	\label{fig:spectra_horizontal}
\end{figure}

Results for the horizontal configuration ($x_1$--$x_3$, parallel to the aerofoil) are addressed in the following.
The eigenvalues spectra are displayed in Figure~\ref{fig:spectra_horizontal} for SPOD-u (left subplot) and for SPOD-p (right subplot).
The spectra are similar to those of the vertical configuration ($x_1$--$x_2$, Figure~\ref{fig:spectra_vertical}), although for the SPOD-u eigenvalues the separation between the rank-1 and 2 modes is much clearer.
Moreover, the rank-1 SPOD-u eigenvalue accounts for approximately 80\% of the flow energy. 
The same hypothesis concerning the rank-1 mode connected to the rod wake/aerofoil interaction for $\mathit{St} \gtrsim 0.3$ and the rank-2 mode for $\mathit{St} \lesssim 0.3$ (aerofoil mode) is made for the SPOD-p spectra and will be further explored with the modes' shapes in the following.

%%%%%%%%%%%%%%%%%%%%%%%%%%%%%%%%%%%%%%%%%%%%%%
\subsubsection{SPOD modes}
\label{sec:results_eigenfunctions}

\begin{figure}[h]
	\centering
	\includegraphics[width=\textwidth]{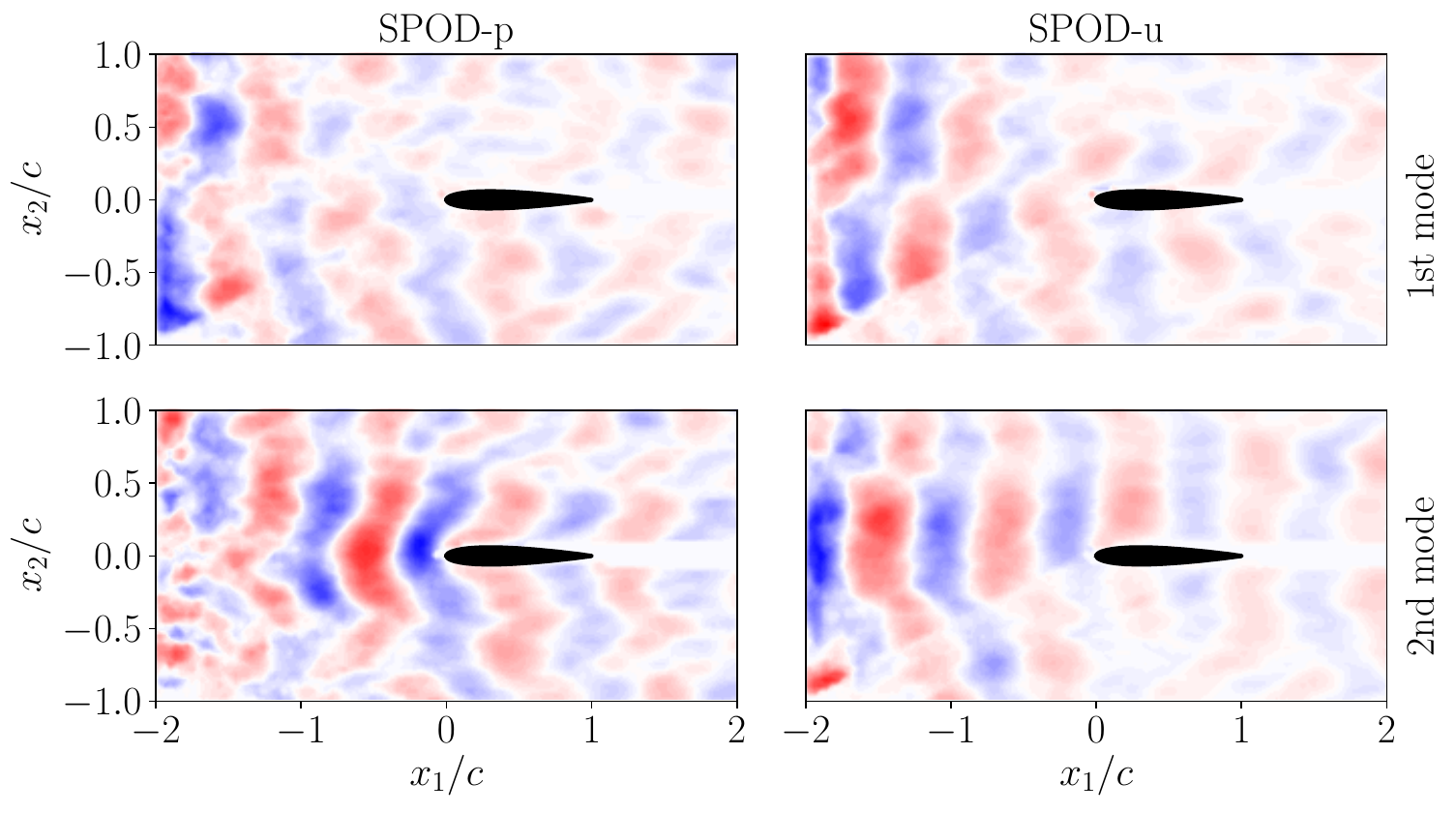}
	\caption{$u_2$-component SPOD modes (real part) for $\mathit{St} \approx 0.2$ and vertical plane configuration ($x_1$--$x_2$). Top and bottom frames denote rank-1 and 2 modes, respectively. Left and right frames indicate SPOD-p and SPOD-u, respectively. The aerofoil is indicated by the black shaded area, whereas the rod is located at $x_1/c = -2$, spanning the entire $x_2$ direction. The red to blue colour map denotes positive to negative values.}
	\label{fig:u2modes_St02_vertical}
	%\modif{ON DIT QUE ROUGE EST POSITIF ET BLEU NEGATIF. C'EST UN PEU IMPRECIS QUANTITATIVEMENT. EST-CE QU'IL NE FAUDRAIT PAS REGARDER L'AMPLITUDE DE LA VITESSE U2 DIVIS\'EE PAR LA NORME DU MODE (QUI INCLUERAIT DONC LES AUTRES COMPOSANTES U1 et U3) AFIN DE VOIR L'IMPORTANCE DE U2 DANS LE MODE CONSIDERE? J'imagine qu'alors la figure en haut a gauche (associee au cylindre, donc avec du u1 et du u3) aurait des couleurs plus pales que la figure en bas a gauche. Peut-etre que le 1er mode de SPOD-u serait plus pale que le rank-2 mode egalement. Cela permettrait de dire que le rank-2 mode est plus organisé (en terme de u2) que le premier comme c'etait ecrit.}
\end{figure}

\begin{figure}[h]
	\centering
	\includegraphics[width=\textwidth]{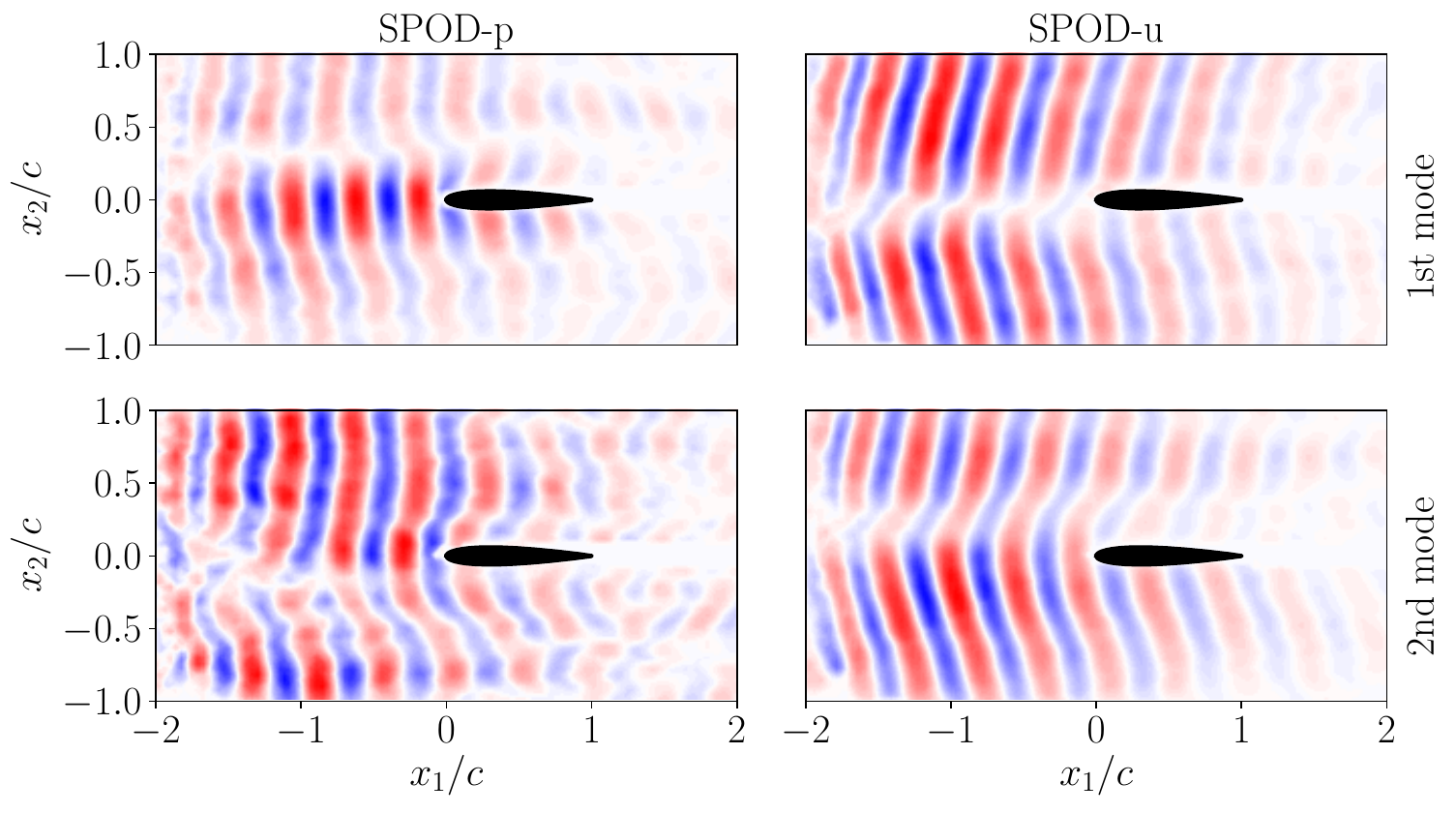}
	\caption{$u_2$-component SPOD modes (real part) for $\mathit{St} \approx 0.4$ and vertical plane configuration ($x_1$--$x_2$). See the comments in the caption of Figure~\ref{fig:u2modes_St02_vertical}.}
	\label{fig:u2modes_St04_vertical}
\end{figure}

Figures~\ref{fig:u2modes_St02_vertical} and \ref{fig:u2modes_St04_vertical} display the rank-1 and 2 SPOD-u and SPOD-p modes for the $u_2$ velocity component (upwash/downwash direction, Figure~\ref{fig:sketch_PIV_acoustics}), obtained in the TR-PIV domain, for $\mathit{St} \approx 0.2$ and 0.4, respectively.
Note that the TR-PIV measurements cannot access an area behind the aerofoil trailing edge, which is marked in white and is a shadow zone owing to the lasers and aerofoil relative position.
The colour map from red to blue indicates positive to negative values.
Some modes also display an inclined shadow that crosses the measurement plane, which is related to a shadow produced by the cylinder (\S~\ref{sec:methodology_set-up}).
Only the $u_2$ component is addressed since this is the velocity component that has been shown to be the most correlated with the rod wake/aerofoil interaction (\S~\ref{sec:results_coherence}).

At $\mathit{St} \approx 0.2$ for the $u_2$ component (Figure~\ref{fig:u2modes_St02_vertical}), the rank-2 mode for SPOD-p is clearly more organised than the rank-1 mode.
The rank-1 mode, being associated with the rod wake contains relatively more $u_1$ and $u_3$ components.
For $\mathit{St} \approx 0.4$, Figure~\ref{fig:u2modes_St04_vertical} shows that the structures identified with the rank-1 SPOD-p modes are more organised upstream of the aerofoil leading edge.
The rank-1 SPOD-p mode at $\mathit{St} \approx 0.4$ shows approximately the same features as the rank-2 SPOD-p mode at $\mathit{St} \approx 0.2$.
The main differences lie in the $x_1$ wavelength, which decreases as the frequency increases, and the fact that, at $\mathit{St} \approx 0.4$, the identified structures are more organised.
This observation reinforces the hypothesis that the rank-2 SPOD-p mode is related to the rod wake/aerofoil interaction for $\mathit{St} \lesssim 0.3$, whereas the rank-1 SPOD-p mode tracks the same structures for $\mathit{St} \gtrsim 0.3$, as indicated by the spectra in Figure~\ref{fig:spectra_vertical} (aerofoil branch).
Comparing the rank-1 SPOD-p and SPOD-u modes in Figure~\ref{fig:u2modes_St04_vertical}, one sees that the SPOD-u mode has a $u_2$ component which is approximately anti-symmetric with respect to the plane $x_2=0$ (aerofoil chord symmetry line).
The aerofoil is located within this nodal plane. %, so that this mode is not producing sound efficiently.
On the contrary, the rank-1 SPOD-p mode %, associated with maximal sound radiation,
corresponds to a $u_2$ component which is approximately symmetric.

\begin{figure}
	\centering
	\includegraphics[width=\textwidth]{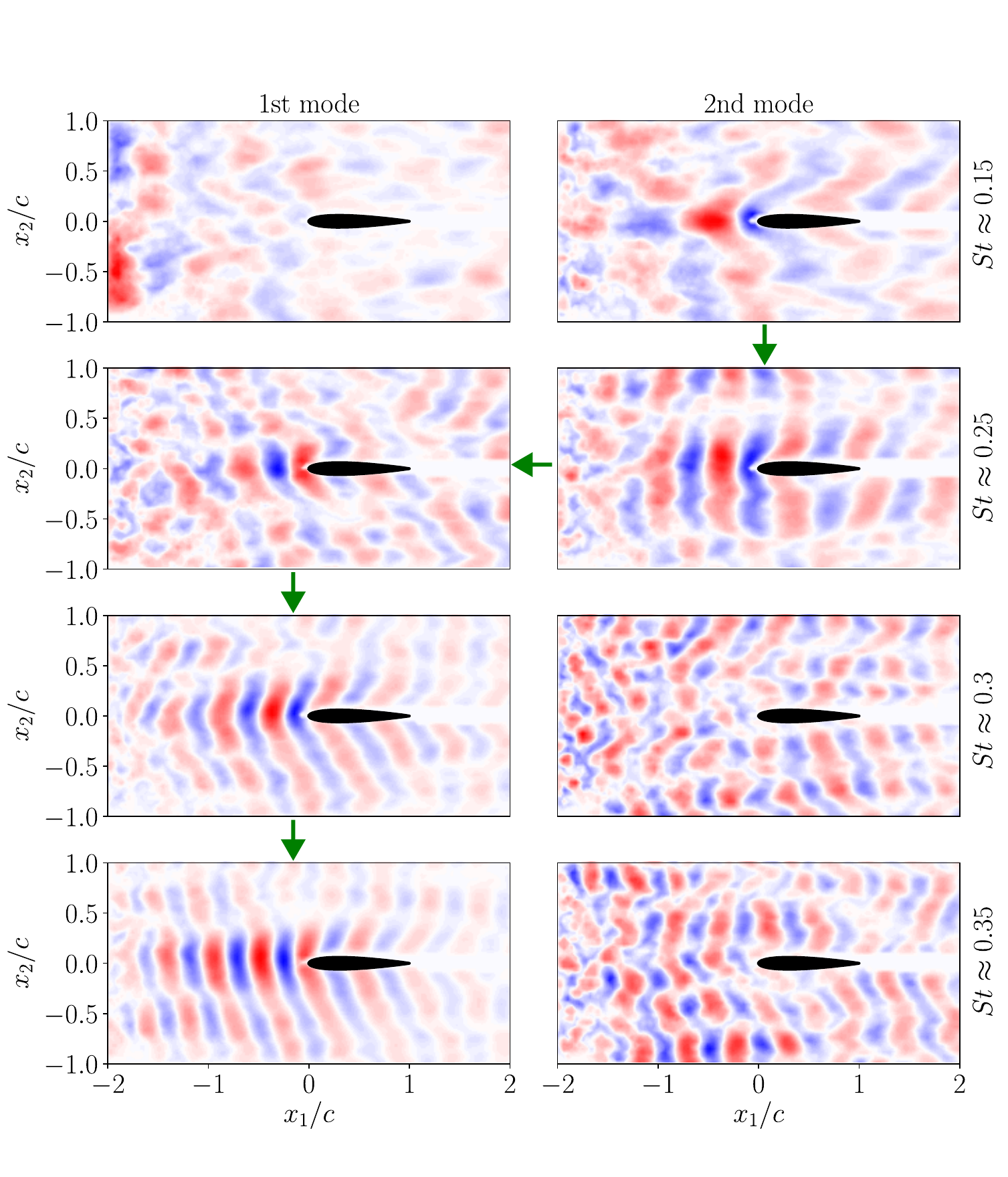}
	\caption{$u_2$-component SPOD-p modes (real part) and vertical plane configuration ($x_1$--$x_2$). Frames, from top to bottom: $\mathit{St} \approx$ 0.15, 0.25, 0.3 and 0.35. The green arrows indicate the mode switching path. Left and right frames indicate the rank-1 and 2 modes, respectively. See the comments in the caption of Figure~\ref{fig:u2modes_St02_vertical}.}
	\label{fig:u2modes_SPODp_vertical}
\end{figure}

In order to track the aerofoil branch under a change of the Strouhal number (see Figures~\ref{fig:spectra_vertical} and~\ref{fig:spectra_horizontal}, the rank-1 and 2 SPOD-p modes for the upwash/downwash velocity component ($u_2$) are shown in Figure~\ref{fig:u2modes_SPODp_vertical} at several Strouhal numbers.
The aerofoil branch is clearly observed in the rank-2 SPOD-p mode at $\mathit{St} \approx 0.15$ and 0.25, as well as in the rank-1 mode at $\mathit{St} \approx 0.30$ and 0.35, consistent with the modes shown in 
Figure~\ref{fig:u2modes_St02_vertical} and the spectra presented in Figure~\ref{fig:spectra_vertical}.
The $u_2$-component structures for the aerofoil branch are more organised, forming a wave-train in the $x_2$-direction, extending from the rod wake to the aerofoil leading edge.
For the cylinder branch, on the other hand, the structures are smaller and less organised, $u_2$ not being the most important component to describe the cylinder wake, compared to $u_1$ and $u_3$.
An idealised two-dimensional von Kármán street generates no velocity fluctuation in the cylinder axis direction ($u_2$).
However, a small $u_2$ component exists as the flow is distorted by the finite cylinder span and the presence and impingement of the flow on the aerofoil leading-edge. 

\begin{figure}
	\centering
	\includegraphics[width=\textwidth]{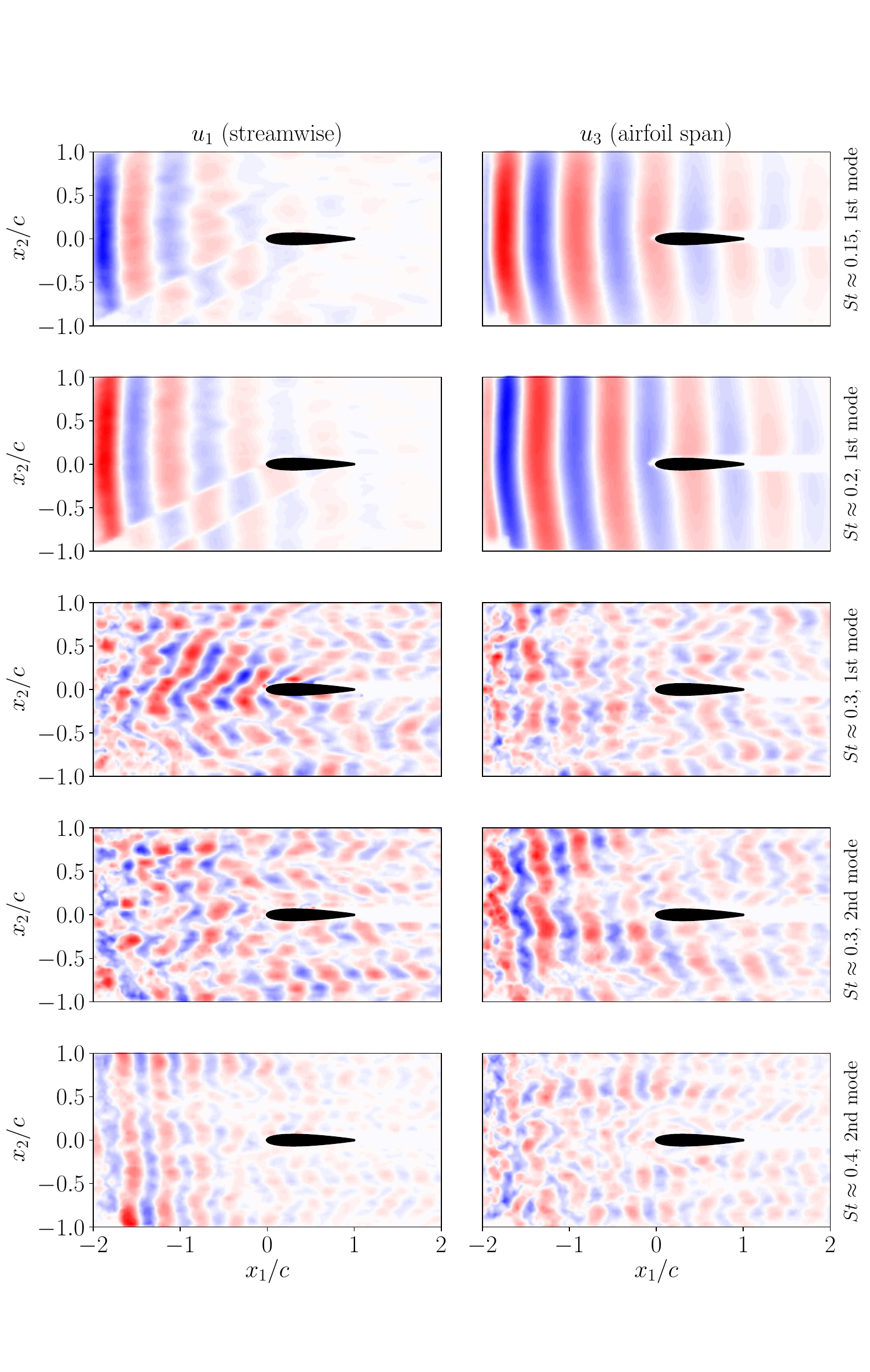}
	\caption{$u_1$- (left frames) and $u_3$-components (right frames) SPOD-p modes (real part) and vertical plane configuration ($x_1$--$x_2$). Frames, from top to bottom: $\mathit{St} \approx$ 0.15, 0.2, 0.3 and 0.4. The three top lines correspond to the rank-1 mode, whereas the two bottom lines correspond to the rank-2 mode. See the comments in the caption of Figure~\ref{fig:u2modes_St02_vertical}.}
	\label{fig:u1u3modes_SPODp_vertical}
\end{figure}

The streamwise ($u_1$) and aerofoil span ($u_3$) velocity SPOD-p modes are displayed in Figure~\ref{fig:u1u3modes_SPODp_vertical} at several Strouhal numbers.
In this figure, the rank-1 and 2 SPOD-p modes are shown for $\mathit{St} < 0.3$ and $\mathit{St} > 0.3$, respectively, to track the cylinder branch.
At $\mathit{St} \approx 0.3$, both the rank-1 and 2 SPOD-p modes are presented.
For $\mathit{St} < 0.3$, the two velocity component modes exhibit highly organised structures that are aligned with the cylinder span ($x_2$ direction).
Note the modes exhibit a certain periodicity regarding the streamwise direction ($x_1$), which corresponds to the wavelength of the von Kármán vortex shedding $\lambda = d/\mathit{St}$, i.e. $\lambda \approx = 0.1$ m when considering $\mathit{St} \approx 0.2$.

At higher frequencies, the structures become less organised while still retaining the alignment pattern.
At the branch-crossing frequency, i.e., $\mathit{St} \approx 0.3$, the $u_3$ velocity component appears to be the most coherent with the von Kármán street, especially in the rank-2 SPOD-p mode, although the $u_1$ velocity component also shows some level of organization, particularly in the rank-1 SPOD-p mode.
At $\mathit{St} \approx 0.4$, the rank-2 SPOD-p mode for the $u_1$ velocity component aligns well with the von Kármán street, and some level of organization is still observed in the $u_3$ velocity component, albeit less organised.

\begin{figure}
	\centering
	\includegraphics[width=\textwidth]{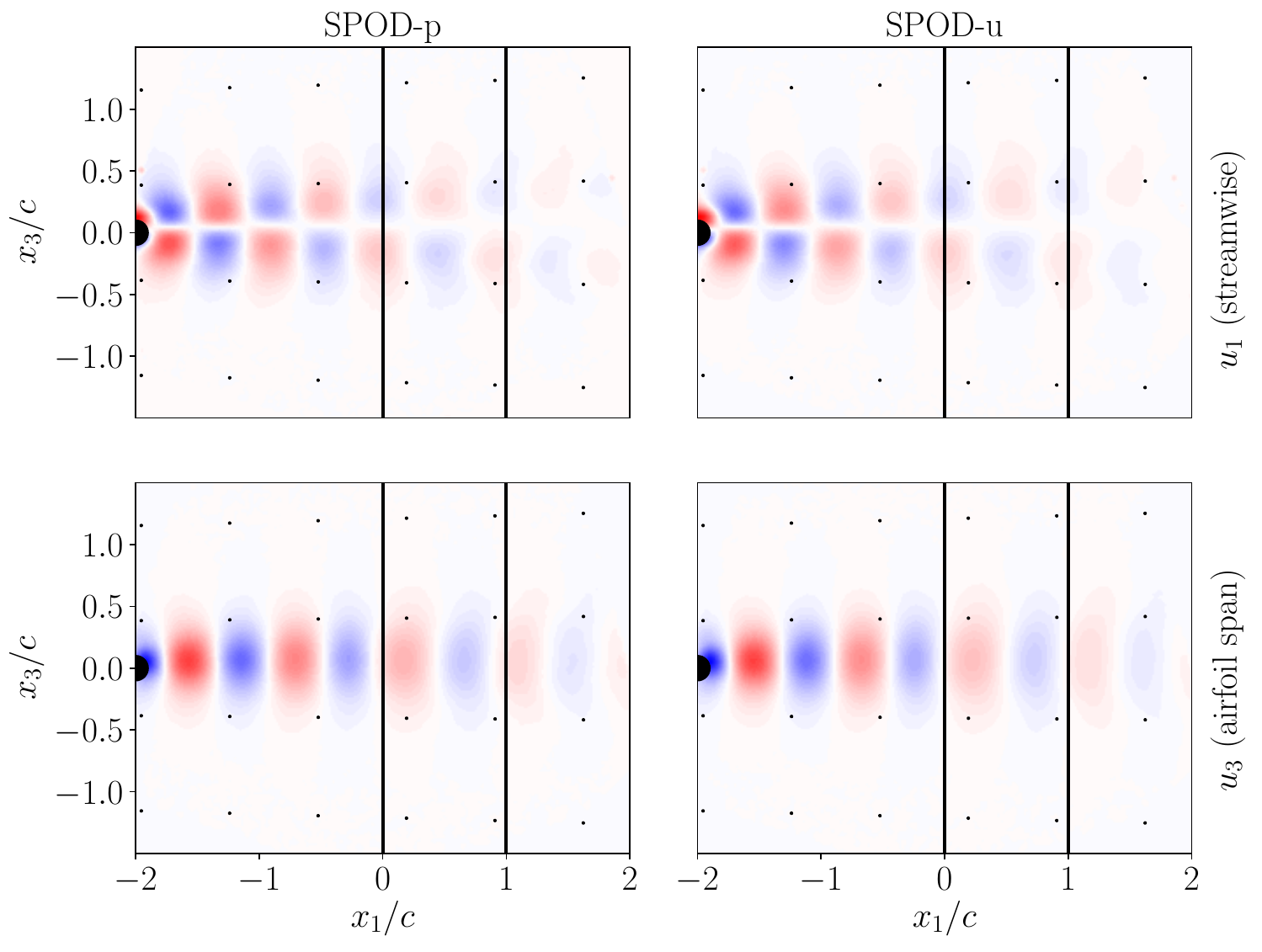}
	\caption{$u_1$ (top frames) and $u_3$ (bottom frames) components' rank-1 SPOD modes (real part) for $\mathit{St} \approx 0.2$ and horizontal plane configuration ($x_1$--$x_3$). Left and right frames indicate SPOD-p and SPOD-u, respectively. Small dots represent the digital microphones array; the closed circle and the two vertical lines indicate the cylinder and the aerofoil, respectively. Red to blue colour map denotes positive to negative values.}
	\label{fig:u13mode_St02_horizontal}
\end{figure}

To be comprehensive, the modes computed for the horizontal plane ($x_1$--$x_3$) are now presented. Figure~\ref{fig:u13mode_St02_horizontal} shows the rank-1 mode for the $u_1$ and $u_3$ velocity components, as well as SPOD-p and SPOD-u for the horizontal configuration and $\mathit{St} \approx 0.2$.
The structures corresponding to the $u_1$ and $u_3$ component are associated with the von Kármán vortex street: the rank-1 SPOD-u and SPOD-p modes at $\mathit{St} \approx 0.2$ are similar, because the most energetic velocity structures in the flow are also those which are associated to the lift dipole radiation aeolian tonal sound at this Strouhal number.
The rank-2 SPOD-p and SPOD-u modes for the $u_1$ and $u_3$ components exhibit similar structures at $\mathit{St} \approx 0.2$ and will not be shown here for brevity.

\begin{figure}
	\centering
	\includegraphics[width=\textwidth]{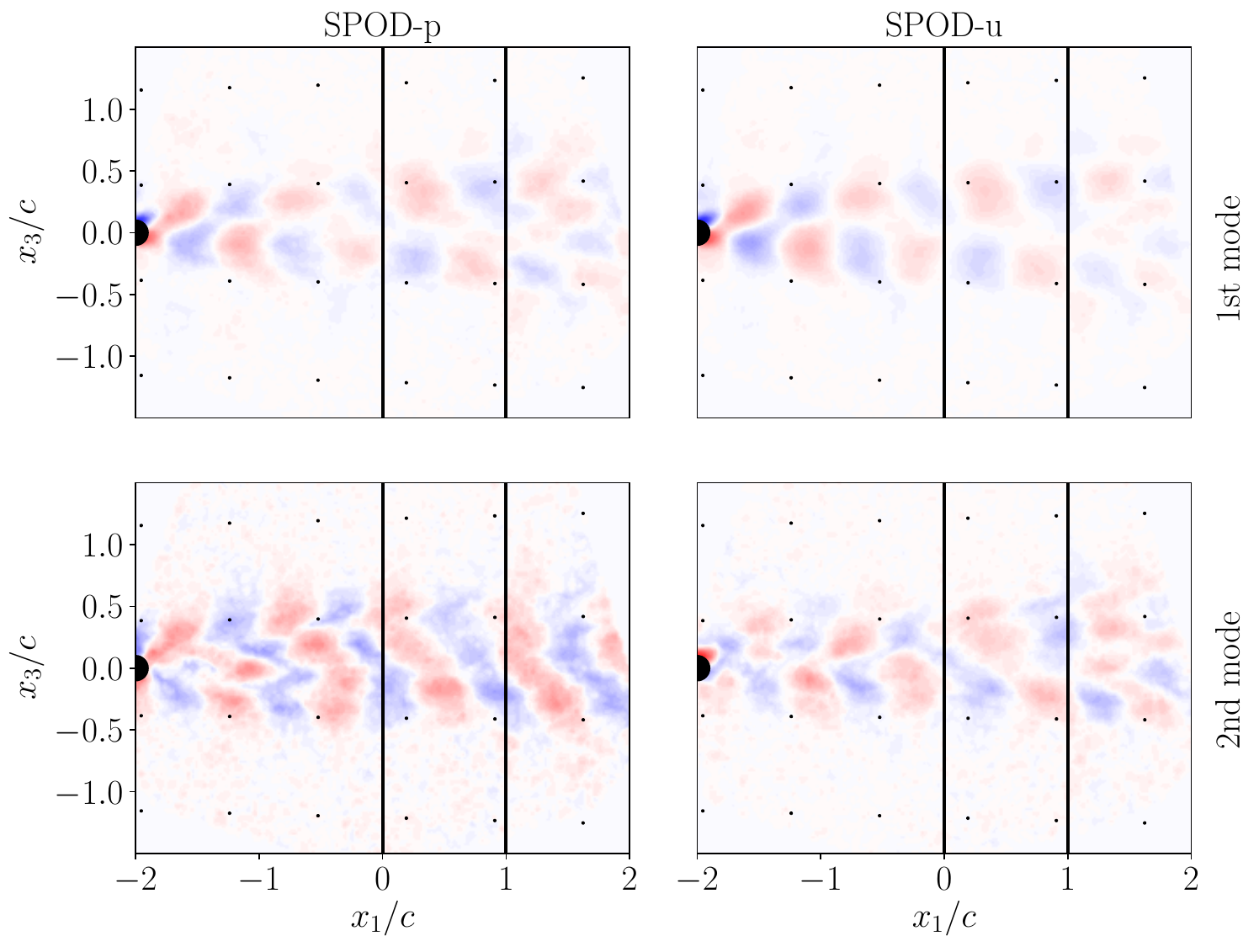}
	\caption{$u_2$-component SPOD-p and SPOD-u modes (real part) for $\mathit{St} \approx 0.2$ and horizontal plane configuration ($x_1$--$x_3$). Top and bottom frames denote rank-1 and 2 modes, respectively. Left and right frames indicate SPOD-p and SPOD-u, respectively. See the comments in the caption of Figure~\ref{fig:u13mode_St02_horizontal}.}
	\label{fig:u2modes_St02_horizontal}
\end{figure}

Regarding the $u_2$ velocity component at $\mathit{St} \approx 0.2$, presented in Figure~\ref{fig:u2modes_St02_horizontal}, the rank-1 SPOD-p and SPOD-u modes seem to be associated with the von Kármán vortex street cylinder branch.
On the other hand, the rank-2 SPOD-p mode is more spread, though less organised, in the aerofoil spanwise ($x_3$) direction.

\begin{figure}
	\centering
	\includegraphics[width=\textwidth]{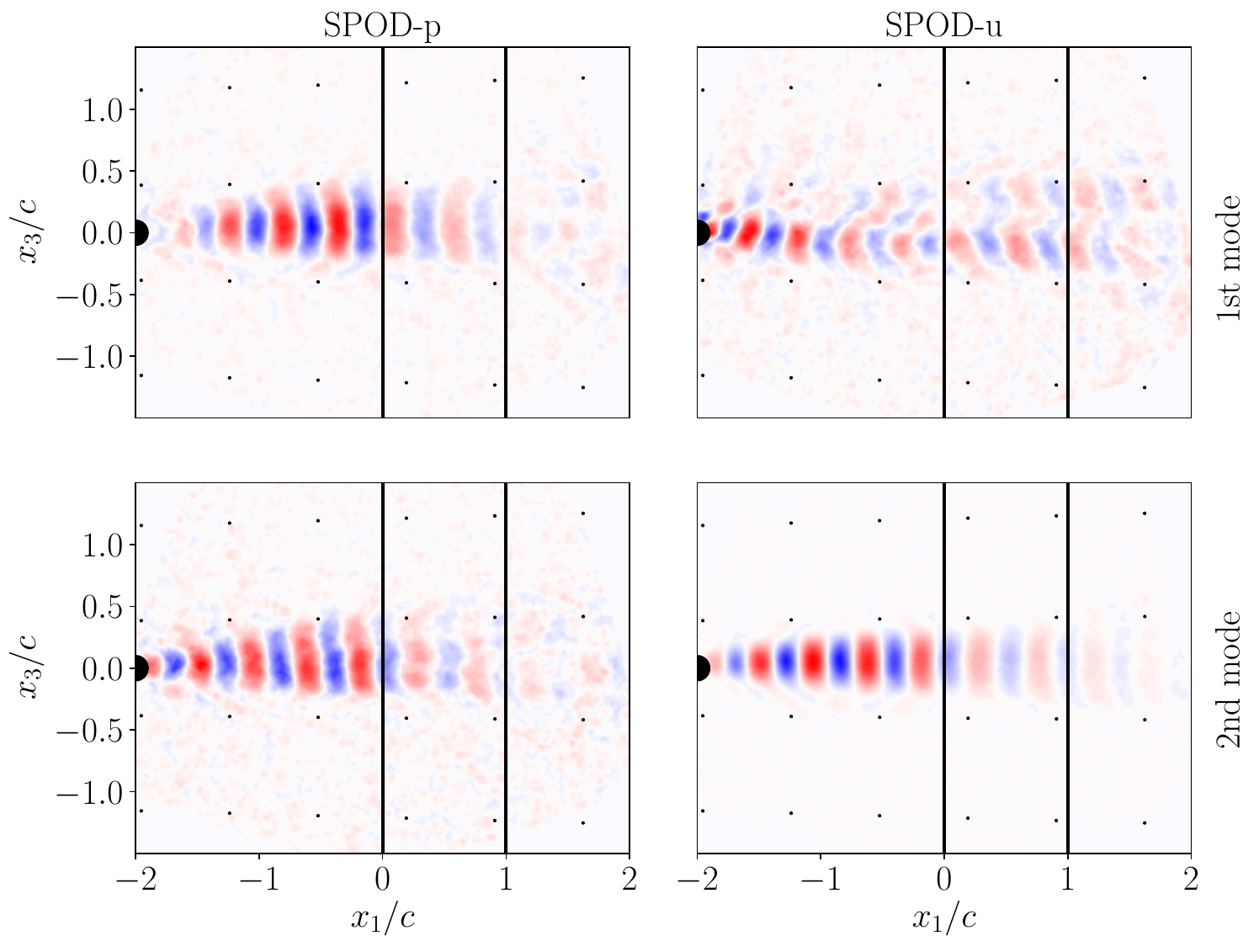}
	\caption{$u_2$-component SPOD-p and SPOD-u modes (real part) for $\mathit{St} \approx 0.4$ and horizontal plane configuration ($x_1$--$x_3$). See the comments in the caption of Figure~\ref{fig:u2modes_St02_horizontal}.}
	\label{fig:u2modes_St04_horizontal}
\end{figure}

The scenario is quite different for $\mathit{St} \approx 0.4$ (Figure~\ref{fig:u2modes_St04_horizontal}), which displays the rank-1 and 2 modes for the $u_2$ velocity component along with SPOD-p and SPOD-u, in the same manner as Figures~\ref{fig:u2modes_St02_vertical} and \ref{fig:u2modes_St04_vertical}, but now for the horizontal configuration.
Wave-train structures emerge in the rod wake.
For SPOD-p, the rank-1 and 2 modes are quite similar and appear to represent rod wake/aerofoil interaction structures.
For SPOD-u, on the other hand, the rank-2 mode is more organised and well-defined as a wave-train structure.
%\modif{JUSTE UNE REMARQUE: le plan horizontal est juste au-dessus du profil, et dans ce plan on ne distingue pas si le profil de u2 est symetrique ou antisymetrique en x2. Memes si les deux modes SPOD-p semble les memes dans ce plan, ce n'etait pas le cas dans un plan vertical. De plus les couleurs rouge/bleu donnent le signe, mais pas le niveau (relatif a toutes les compostantes). Donc si ca se trouve, le niveau de u2 relatif est comparativement plus grand dans le 2nd mode SPOD-p que dans le premier (le plan horizontal etant proche du plan d'antisymmetrie dans ce dernier cas).}
One should note the horizontal plane ($x_1$--$x_3$) is located just above the aerofoil (see Figure~\ref{fig:sketch_PIV_acoustics}), which makes it impossible to distinguish if the $u_2$ structures are symmetric or antisymmetric regarding the $x_2$ direction.

%Acoustic pressure modes for the horizontal configuration at $\mathit{St} \approx 0.2$ and 0.4 are similar to those obtained with the vertical configuration, shown in Figures~\ref{fig:pmodes_St02_vertical} and \ref{fig:pmodes_St04_vertical}, respectively, and therefore will not be shown here for brevity.
%The same is can be said about the modes convergence, with the horizontal and vertical configurations displaying very similar levels; hence results for the former and will not be addressed for brevity.

%%%%%%%%%%%%%%%%%%%%%%%%%%%%%%%%%%%%%%%%%%%%%%
\subsection{Beamforming maps}
\label{sec:results_beamforming}

\begin{figure}[h!]
	\centering
	\includegraphics[width=\textwidth]{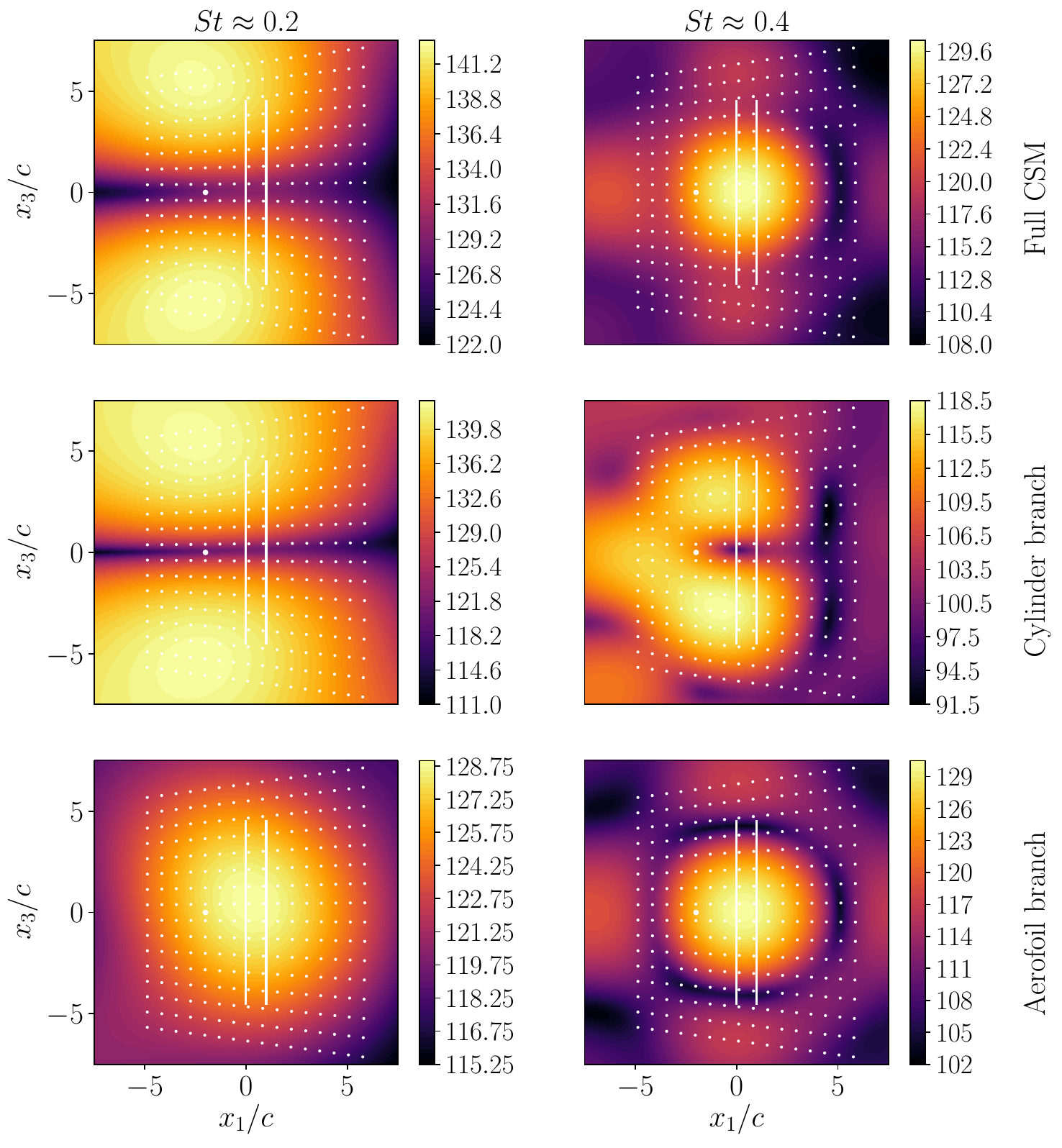}
	\caption{Beamforming maps for the vertical configuration ($x_1$--$x_2$) at $\mathit{St} \approx 0.2$ (left panels) and $\mathit{St} \approx 0.4$ (right panels). Results are shown for the full CSM (top panels), cylinder branch (middle panels), and aerofoil branch (bottom panels). Acoustic levels are presented in dB/St. The microphone array (dots), rod (filled circle), and aerofoil (solid lines) are also indicated.}
	\label{fig:beamforming_St02_04_vertical}
\end{figure}

%In the previous section, we have deduced reasonable origins for the dipole sources (either rod or aerofoil) based on the position of the lobes observed in the pressure amplitude maps (figures~\ref{fig:pmodes_St02_vertical}-\ref{fig:pmodes_St04_vertical}).
In order to complement the analysis, we use the beamforming technique to localize sources associated with given pressure SPOD modes.
Figure~\ref{fig:beamforming_St02_04_vertical} presents beamforming maps for $\mathit{St} \approx 0.2$ and 0.4, obtained using the full microphone array CSM~\eqref{eq:CSM} and SPOD-p low-rank models~\eqref{eq:CSM_low-rank} associated with the aerofoil and cylinder branches.
Only results for the vertical configuration ($x_1$--$x_2$) are shown, as those for the horizontal configuration ($x_1$--$x_3$) are qualitatively similar and omitted for brevity.
The acoustic map levels are given in dB/St, and projections of the microphone array, rod, and aerofoil onto the beamforming maps are also displayed.

When beamforming is performed using the full CSM~\eqref{eq:beamforming}, top panels of Figure~\ref{fig:beamforming_St02_04_vertical}), the acoustic source at $\mathit{St} \approx 0.2$ appears as a dipole aligned with the rod spanwise direction, with maximum power levels located around $(x_1/c, x_3/c) \approx (-3.5, \pm5)$.
It should be noted that at such a low frequency, the array resolution is limited -- approximately 1.1~m (see Appendix~\ref{app:bw_dr}, Figure~\ref{fig:bw_dr}).
At $\mathit{St} \approx 0.4$, the source is mapped centred on the aerofoil leading-edge, with peak levels located at the aerofoil's leading-edge mid-span.
At this frequency, the array resolution improves to approximately 0.55~m.

For the cylinder branch low-rank model, the CSM~\eqref{eq:CSM_low-rank} and beamforming ~\eqref{eq:beamforming_low-rank} are computed using the rank-1 SPOD-p mode for $\mathit{St} < 0.3$ and the rank-2 SPOD-p mode for $\mathit{St} \geq 0.3$.
Conversely, for the aerofoil branch, the rank-2 SPOD-p mode is used for $\mathit{St} < 0.3$, and the rank-1 mode for $\mathit{St} \geq 0.3$.
These mode selections are based on the identification of cylinder and aerofoil branches in the SPOD-p eigenvalue spectra (figures~\ref{fig:spectra_vertical} and \ref{fig:spectra_horizontal}).
The same branches could not be clearly identified using SPOD-u, whose modes yielded beamforming maps closely resembling those of the full CSM; thus, SPOD-u results will not be further addressed here.

Beamforming maps based on the cylinder branch (figure \ref{fig:beamforming_St02_04_vertical} middle panels) show dipole-like lobes at both $\mathit{St} \approx 0.2$ and 0.4, aligned with the rod span ($x_3$-direction).
When conventional beamforming is used--which models sources as uncorrelated, compact monopoles--dipole sources appear as two symmetric energy spots with respect to the source location (see Figure 1 in~\citet{porteous2015three}).
Due to array resolution limitations (Appendix~\ref{app:bw_dr}), the lobes are broader at $\mathit{St} \approx 0.2$ than at $\mathit{St} \approx 0.4$.
At $\mathit{St} \approx 0.4$, the peak levels are located around $(x_1/c, x_3/c) \approx (-2, \pm2.5)$.
The aerofoil branch maps (figure \ref{fig:beamforming_St02_04_vertical} bottom panels) reveal sources at the aerofoil leading-edge at both frequencies, with peak levels again near the aerofoil leading-edge mid-span.
As expected from the improved resolution at higher frequency, the source appears slightly more compact at $\mathit{St} \approx 0.4$.

These results support the hypothesis of distinct cylinder and aerofoil branches discussed in \S~\ref{sec:results_structures}.
Conventional beamforming identifies sources that align with expectations -- dipole-like structures associated with the rod/von Kármán street and a more compact source near the aerofoil leading edge associated with rod--aerofoil interaction.

Further improvements in source localization could be achieved by employing a dipole Green's function formulation~\eqref{eq:green_function}, especially for sources attributed to the cylinder branch.
Additionally, deconvolution algorithms such as DAMAS~\citep{brooks2006damas} and CLEAN-SC~\citep{sijtsma2007clean} could mitigate the limited resolution of the microphones array.
However, such enhancements -- along with beamforming analysis of other frequencies and low-rank models aimed at identifying additional physical mechanisms -- are beyond the scope of this work, which focuses on characterizing the rod--aerofoil interaction sound--source.

%%%%%%%%%%%%%%%%%%%%%%%%%%%%%%%%%%%%%%%%%%%%%%
\subsection{Comparison between acoustic and SPOD spectra}
\label{sec:results_spectra}

\begin{figure}
	\centering
	\includegraphics[width=0.6\textwidth]{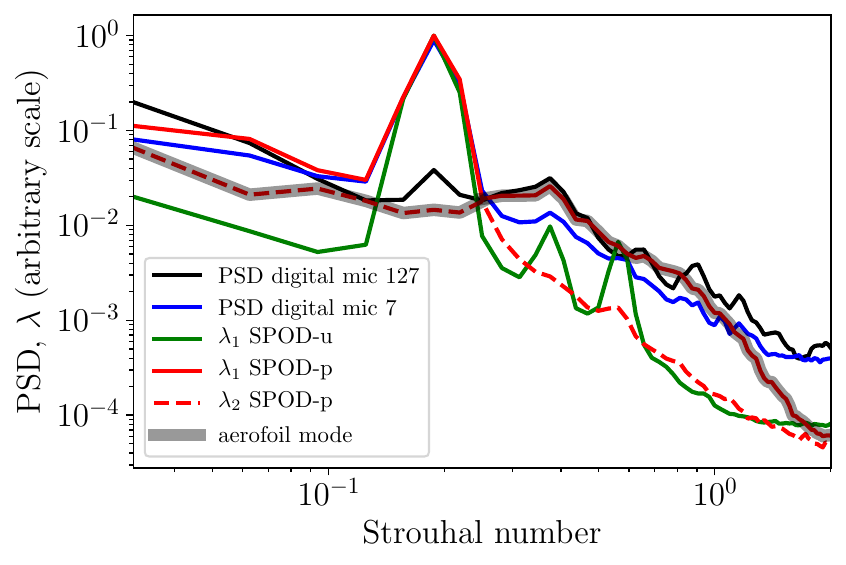}
	\caption{Acoustic and SPOD/ESPOD spectra for the vertical configuration ($x_1$--$x_2$).}
	\label{fig:acoustic_spod_spectra}
\end{figure}

In this section a qualitative comparison among the SPOD and acoustic spectra is discussed.
Figure~\ref{fig:acoustic_spod_spectra} exhibits the acoustic spectra for digital microphones \#127 (located above the aerofoil midspan and mid-chord, black curve) and \#7 (positioned above the aerofoil side, blue curve), as indicated  in Figure~\ref{fig:acoustic_microphones}, along with the eigenvalue spectra for SPOD-u (green curve) and SPOD-p (red curves, with continuous and dashed lines denoting the rank-1 and 2 modes, respectively).
To enable a proper comparison, the intensity scale in the figure is arbitrary, with normalized values such that their maximum is one.
Only results for the TR-PIV vertical configuration ($x_1$--$x_2$, Figure~\ref{fig:sketch_PIV_acoustics}) are shown for brevity, but similar spectra were obtained for the horizontal configuration ($x_1$--$x_3$).

The broad peak at the vortex shedding frequency, $\mathit{St} \approx 0.2$, is clearly visible in the spectra obtained with the microphone on the aerofoil side (mic \#7, blue curve) and for the rank-1 mode of both SPOD-u and SPOD-p spectra (continuous green and red curves).
The interaction noise peak is more evident in the spectra measured with the microphone above the aerofoil midspan/mid-chord (mic \#127, black curve) and for the rank-1 mode of both the SPOD-u and SPOD-p spectra (continuous green and red curves).
Note that, up to $\mathit{St} \approx 0.3$, the eigenspectrum of the rank-2 mode of SPOD-p (dashed red curve) does not contain the aeolian tonal, in the same fashion as the aerofoil midspan/mid-chord (mic \#127, black curve).
At the $0.3 \lesssim \mathit{St} \lesssim 1$ range, on the other hand, the acoustic spectrum is more similar to the rank-1 SPOD-p eigenspectrum.
The mode switching is indicated in the figure by the grey thick curve and is denoted as \emph{aerofoil mode}, as was done in Figures~\ref{fig:spectra_vertical} and \ref{fig:spectra_horizontal} using a blue thick curve instead of a grey one.
The resemblance between the acoustic spectrum of the microphone above the aerofoil and the proposed aerofoil mode provides further support for the hypothesis of mode switching for Strouhal numbers lower and higher than 0.3, together with the structures identified by SPOD-p in \S~\ref{sec:results_structures}.

%%%%%%%%%%%%%%%%%%%%%%%%%%%%%%%%%%%%%%%%%%%%%%
\section{Conclusions}
\label{sec:conclusions}
%%%%%%%%%%%%%%%%%%%%%%%%%%%%%%%%%%%%%%%%%%%%%%

In this paper, a perpendicular rod/aerofoil configuration was studied using TR-PIV and microphone array measurements.  
The data were simultaneously acquired with both instruments, allowing for the computation of correlations between the velocity and acoustic fields.  
The TR-PIV experiments were conducted in two measurement planes: one aligned with the rod/cylinder spanwise direction (lateral/vertical plane, $x_1$--$x_2$) and the other aligned with the aerofoil spanwise direction (top/horizontal plane, $x_1$--$x_3$).  
SPOD modal decompositions were also performed to characterize the coherent structures present in the flow. Modal decompositions were obtained using inner products targeting either pressure (SPOD-p) or turbulence kinetic energy (SPOD-u).
The resulting modes are equivalent to ESPOD modes, as shown in Appendix~\ref{app:methodology_ESPOD}.

Acoustic measurements from the microphone array show broad peaks at $\mathit{St} \approx 0.2$ (vortex shedding frequency) and its harmonics, especially the first harmonic ($\mathit{St} \approx 0.4$).  
Microphones aligned with the rod wake -- positioned above the aerofoil midspan and spanning from upstream of the cylinder to downstream of the aerofoil trailing edge -- measured peaks at $\mathit{St} \approx 0.2$ and $0.4$ with similar intensity levels.  
In contrast, microphones located near the aerofoil tips, oriented perpendicularly to the von Kármán vortex street, recorded a strong peak at $\mathit{St} \approx 0.2$, approximately 15 dB higher than the first harmonic peak, linked to a dipole perpendicular to the von Kármán vortex street.  
These results are further reinforced by conventional beamforming maps.

The coherence between a microphone in the acoustic field and the three velocity components reached levels of up to 60\% when considering the $u_2$ (upwash/downwash direction) velocity component at $\mathit{St} \approx 0.4$ and the vertical plane ($x_1$--$x_2$).  
For the horizontal plane ($x_1$--$x_3$), such levels were as high as 40\%, again for the $u_2$ component.  
Regarding the spatial region, the highest coherence levels occurred in the rod wake, near the position where the flow impinges on the aerofoil leading edge.  
These findings corroborate previous literature concerning the contribution of the upwash/downwash velocity component to the sound source~\citep{howe1988contributions} and the aerofoil leading edge as the source region~\citep{jacob2005rod, boudet2005wake}.  
It is important to emphasize that the present study identified $\mathit{St} \approx 0.4$, i.e., the first harmonic of the vortex shedding frequency, which is linked to the cylinder drag~\citep{giret2015noise}, as the frequency at which the rod/aerofoil interaction noise is most evident.  
For completeness, at $\mathit{St} \approx 0.2$, i.e., the vortex shedding frequency, the highest coherence levels, reaching up to 20\%, pertain to the $u_1$ (streamwise direction) and $u_3$ (aerofoil span direction) velocity components, which are associated with the von Kármán vortex street.

SPOD analysis corroborates the coherence results, showing that the structures of the $u_2$ (upwash/downwash) velocity component are more organised in the rod wake and impinge upon the aerofoil leading edge.  
This is especially true when the SPOD is calculated using an inner product targeting the pressure/acoustic component (SPOD-p) at $\mathit{St} \approx 0.4$ for both TR-PIV planes, which, overall, presents higher convergence levels.  
In these cases, the eigenspectra exhibit low-rank characteristics.  
Interestingly, a similar structure is observed at $\mathit{St} \approx 0.2$ in the vertical TR-PIV plane, but this time as the rank-2 SPOD-p mode.  
Upon further analysis of the eigenvalue spectra, it is observed that up to $\mathit{St} \approx 0.3$, the rank-2 mode seems to be related to the rod/aerofoil interaction noise, whereas for frequencies higher than $\mathit{St} \approx 0.3$, the rank-1 mode is responsible for such interaction.  
Regarding the von Kármán structures, the rank-1 SPOD-p mode up to $\mathit{St} \approx 0.3$ and the rank-2 SPOD-p mode for frequencies higher than $\mathit{St} \approx 0.3$ are linked to such mechanism, with a dipole perpendicular to the von Kármán vortex street being identified.  
A similar mode-switching behaviour has been previously observed in jet noise problems~\citep{schmidt2018spectral, lesshafft2019resolvent}.  
Beamforming maps obtained using low-rank models of the microphones--cross-spectral matrix (CSM) further reinforce the aerofoil and cylinder modes/branches hypothesis.

The present study underscores the importance of careful selection of a strategy to extract coherent structures linked to sound generation.  
When dealing with databases that contain the velocity field, as in TR-PIV experiments, the obvious choice of a TKE inner product may not always identify the appropriate structures.  
If simultaneous acoustic measurements are available, an inner product targeting the acoustic field may yield structures that are more coherent with the problem at hand, specifically rod/aerofoil interaction.
%\modif{(Est-ce qu'il pourrait etre interessant de redire que les structures energetiques dans l'ecoulement a St=0.4 ont un u2 antisymetrique par rapport au plan de l'aerofoil, alors que les structures correles au son sont symetriques ?)}  
This approach is particularly valuable in noise source modelling efforts, where a proper low-order basis is crucial to provide insights into the physical mechanisms underpinning sound generation.
Ongoing efforts are being made in this direction.

%%%%%%%%%%%%%%%%%%%%%%%%%%%%%%%%%%%%%%%%%%%%%%
\bmhead{Acknowledgements}
%%%%%%%%%%%%%%%%%%%%%%%%%%%%%%%%%%%%%%%%%%%%%%

The authors would like to thank Laurent Philippon, Damien Eysseric, Pascal Biais, Jean-Christophe Vergez and Janick Lamounier for their technical support.
We are also grateful to Vincent Jaunet and L. Bega.

%%%%%%%%%%%%%%%%%%%%%%%%%%%%%%%%%%%%%%%%%%%%%%
\section*{Declarations}
%%%%%%%%%%%%%%%%%%%%%%%%%%%%%%%%%%%%%%%%%%%%%%

\begin{itemize}
	\item This work was supported by DGAC (Direction Générale de l'Aviation Civile), PNRR (Plan National de Relance et de Résilience Français) and NextGeneration EU via project MAMBO (Méthodes Avancés pour la Modélisation du Bruit moteur et aviOn).
	\item The authors report no conflict of interest.
\end{itemize}

%%%%%%%%%%%%%%%%%%%%%%%%%%%%%%%%%%%%%%%%%%%%%%
\begin{appendices}
%%%%%%%%%%%%%%%%%%%%%%%%%%%%%%%%%%%%%%%%%%%%%%

%%%%%%%%%%%%%%%%%%%%%%%%%%%%%%%%%%%%%%%%%%%%%%
\section{Connection between SPOD and ESPOD}
\label{app:methodology_ESPOD}

It is possible to correlate the SPOD modes evaluated for a specific spatial region or flow quantity with another target field.
In the space-time domain, this technique is known as extended proper orthogonal decomposition (EPOD)~\citep{boree2003extended} and can also be adapted to the frequency domain, referred to as extended spectral proper orthogonal decomposition (ESPOD)~\citep{hoarau2006analysis, karban2022self, padillamontero2024eduction}.
ESPOD enables the extraction of, for example, the portion of velocity that is correlated with the pressure expansion coefficients.
The same kind of objectives subtends the use of weighting matrices~\eqref{eq:spod_velocity_weights} or~\eqref{eq:spod_pressure_weights} in SPOD-u and SPOD-p.
In this appendix we indeed show the equivalence between the SPOD-p/SPOD-u modes and ESPOD modes.
Note that, in principle, instead of velocity components $u$ and acoustic component $p$, any other scalar quantity could be employed.

In SPOD-p, we solve the first equation of the system~\eqref{eq:spod_snapshots} using the weight matrix given by~\eqref{eq:spod_pressure_weights}.
We write $\boldsymbol{\hat{Q}} = [\boldsymbol{\hat{Q}}_{u} \; \boldsymbol{\hat{Q}}_{p}]^T$, where, for a given frequency, $\boldsymbol{\hat{Q}}_{p}$ contains the Fourier coefficients for the pressure (obtained using $\boldsymbol{\hat{q}} = \left[\hat{p}\right]^{T}$ in~\eqref{eq:q_matrix}), and $\boldsymbol{\hat{Q}}_{u}$ contains the Fourier coefficients for the velocity (obtained using $\boldsymbol{\hat{q}} = \left[\hat{u}\right]^{T}$ in~\eqref{eq:q_matrix}).
Remember that the pressure and velocity snapshots were acquired in a synchronized manner.
Both of them have $N_b$ columns corresponding to $N_b$ blocks.
Then, performing matrix block multiplication in~\eqref{eq:spod_snapshots} accounting for~\eqref{eq:spod_pressure_weights}, we see that the pressure component is decoupled from the velocity one as
\begin{equation}
    \boldsymbol{\hat{Q}}_p^{\dagger} \boldsymbol{K}_p \boldsymbol{\hat{Q}}_p \boldsymbol{\Psi} =  \boldsymbol{\Psi} \boldsymbol{\Lambda} \mbox{.}
    \label{eq:SPODp_first}
\end{equation}

This is the same SPOD problem that would be solved if only the pressure had been measured.
The solution of this eigenvalue problem provides the expansion coefficients $\boldsymbol{\Psi}_p$ and the eigenvalues $\boldsymbol{\Lambda}_p$, with the subscript $p$ highlighting that they are obtained from pressure only.
From the expansion coefficients, we recover the SPOD-p modes by using the second equation of the system~\eqref{eq:spod_snapshots}, i.e.
\begin{equation}
    \boldsymbol{\Phi} = \left[\begin{array}{cc} \boldsymbol{\Phi}_{u\vert p} \\ \boldsymbol{\Phi}_p \end{array}\right] = \left[\begin{array}{cc} \boldsymbol{\hat{Q}}_u \boldsymbol{\Psi}_p \boldsymbol{\Lambda}_p^{-1/2}\\ \boldsymbol{\hat{Q}}_p \boldsymbol{\Psi}_p \boldsymbol{\Lambda}_p^{-1/2}\end{array}\right] \mbox{.}
    \label{eq:SPODp_second}
\end{equation}

The pressure component of the SPOD-p mode, $\boldsymbol{\Phi}_p = \boldsymbol{\hat{Q}}_p \boldsymbol{\Psi}_p \boldsymbol{\Lambda}_p^{-1/2}$ are the SPOD modes computed from pressure data. On the other hand, the velocity components of the SPOD-p mode given by
\begin{equation}
    \boldsymbol{\Phi}_{u\vert p} = \boldsymbol{\hat{Q}}_u \boldsymbol{\Psi}_p \boldsymbol{\Lambda}_p^{-1/2} \; \mbox{,}
    \label{eq:SPODp_ucomponent}
\end{equation}
\noindent are obtained by combining the velocity snapshots in $\boldsymbol{\hat{Q}}_u$ with pressure expansion coefficients, $\boldsymbol{\Psi}_p \boldsymbol{\Lambda}_p^{-1/2}$.
Thus, the velocity component of the SPOD-p modes is subordinate to the  pressure component, and for this reason are denoted by $\boldsymbol{\Phi}_{u\vert p}$ (rather than by $\boldsymbol{\Phi}_u$).

Doing exactly same analysis with the SPOD-u modes, but using the weight matrix in~\eqref{eq:spod_velocity_weights}, we find that an eigenvalue problem is solved for the velocity data only, which provide SPOD modes having a velocity component $\boldsymbol{\Phi}_u = \boldsymbol{\hat{Q}}_u \boldsymbol{\Psi}_u \boldsymbol{\Lambda}_u^{-1/2}$ and a pressure component
\begin{equation}
   \boldsymbol{\Phi}_{p \vert u} = \boldsymbol{\hat{Q}}_p \boldsymbol{\Psi}_u \boldsymbol{\Lambda}_u^{-1/2} \; \mbox{.}
   \label{eq:SPODu_pcomponent}
\end{equation}
The pressure component is now subordinate to the velocity one.

It turns out that the velocity component of SPOD-p modes~\eqref{eq:SPODp_ucomponent}, and the pressure component of SPOD-u modes~\eqref{eq:SPODu_pcomponent}, correspond exactly to the definition of ESPOD modes~\citep{karban2022self, padillamontero2024eduction}.
Thus using special weight matrices is another way to compute theses modes.

%%%%%%%%%%%%%%%%%%%%%%%%%%%%%%%%%%%%%%%%%%%%%%
\section{Microphones array characteristics\label{app:bw_dr}}

The performance of the microphone array can be characterized by its point spread function (PSF), denoted by $F$.
The PSF represents the response of the array to a unit-amplitude point monopole source positioned at the centre of the source domain~\citep{amaral2018design}, and is defined as
\begin{equation}
	F(\boldsymbol{r}_{o,n},\omega) = \left|\frac{\boldsymbol{g}^\dagger(\boldsymbol{r}_{m,n},\omega) \boldsymbol{g}(\boldsymbol{r}_{m,t},\omega)}{||\boldsymbol{g}(\boldsymbol{r}_{m,n},\omega)||~||\boldsymbol{g}(\boldsymbol{r}_{m,t},\omega)||}\right|^2 \mbox{,}
	\label{eq:psf}
\end{equation}
\noindent where the index $t$ refers to the central position of the scanning grid, i.e., $({x_1}_t,{x_2}_t,{x_3}_t) = (0,0,0)$.

Based on the computed PSF, two key performance metrics of the antenna can be assessed: the beamwidth (or resolution) and the dynamic range.
The beamwidth is defined as the diameter of the PSF main lobe at the level 3 dB below its peak.
The dynamic range corresponds to the difference, in decibels, between the maximum of the main lobe and the highest secondary lobe.

\begin{figure}[h]
	\centering
	\includegraphics[width=0.65\textwidth]{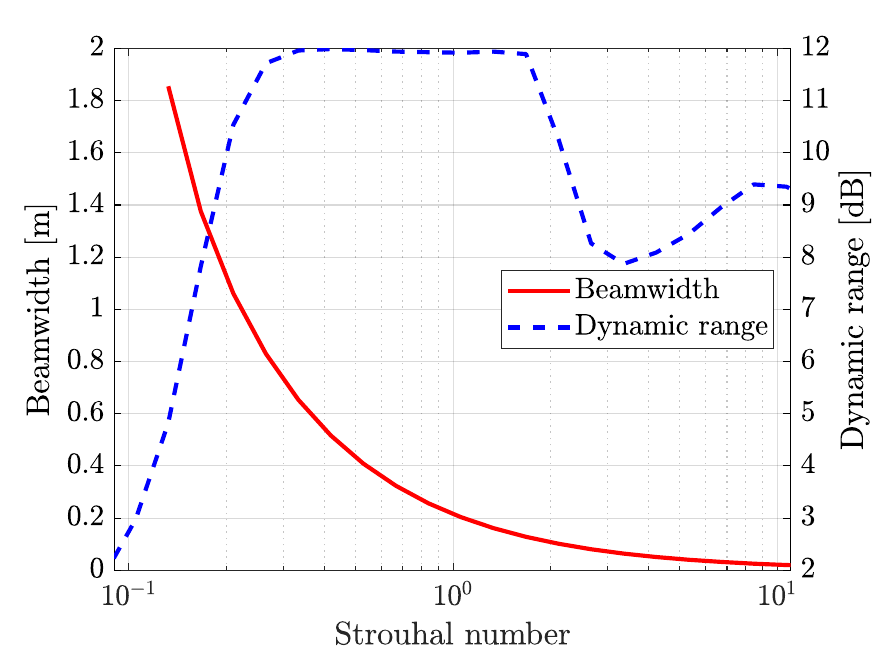}
	\caption{Beamwidth (solid line) and dynamic range (dashed line) of MEMS array B.}
	\label{fig:bw_dr}
\end{figure}

Figure~\ref{fig:bw_dr} presents both metrics for MEMS array B~\citep{zhou2020design},~\citep{zhou2020three} employed in the present study.
The PSF was computed over a scanning grid centred at the mid-span of the aerofoil leading edge, consistent with the beamforming maps, but with extended spatial dimensions of $100c$ (10 m) in both the $x_1$ and $x_2$ directions, as opposed to $15c$ (1.5 m) used elsewhere in this study.

At Strouhal numbers of approximately $\mathit{St} \approx 0.2$ and $0.4$, the array beamwidth is about 1.1~m and 0.55~m, respectively.
For $\mathit{St} < 0.15$, the beamwidth becomes excessively large and exceeds the domain limits of the PSF mesh, rendering its evaluation unreliable under the present configuration.

Regarding the dynamic range, values stabilize around 12 dB in the range $0.3 \lesssim \mathit{St} \lesssim 1.5$, decrease to approximately 8 dB at $\mathit{St} \approx 3.5$, and gradually increase up to 9.5 dB by $\mathit{St} = 11$.
For $\mathit{St} < 0.3$, the secondary PSF peaks (side lobes) extend beyond the mesh boundaries due to the large beamwidth, compromising the reliability of the estimated dynamic range within this frequency regime.

\end{appendices}

%\FloatBarrier

%%%%%%%%%%%%%%%%%%%%%%%%%%%%%%%%%%%%%%%%%%%%%%
%\section{References}
%%%%%%%%%%%%%%%%%%%%%%%%%%%%%%%%%%%%%%%%%%%%%%

\bibliographystyle{sn-mathphys-ay}
\bibliography{references}% common bib file

\end{document}